\documentclass[11pt,a4paper]{article}
\pdfoutput=1

\usepackage{a4wide}
\usepackage{graphicx} 
\usepackage{amsmath}
\usepackage{amssymb}
\usepackage{amsfonts}
\usepackage{cite}
\usepackage{color}
\usepackage{adjustbox}
\usepackage{pdflscape}
\usepackage{gensymb}
\usepackage{graphicx}
\usepackage{diagbox}
\usepackage{pifont}
\usepackage{bigstrut}
\usepackage[small,bf]{caption}
\usepackage{enumerate}
\usepackage{tabulary}
\usepackage{hyperref}
\usepackage{makecell}
\usepackage{multirow}
\usepackage{dcolumn}
\usepackage{textcomp}

\usepackage{ae} 
\usepackage{slashed}
\usepackage{aecompl}
\usepackage{bm}

\usepackage{epsfig}
\usepackage{fontenc}
\usepackage{graphicx}
\usepackage[lofdepth,lotdepth,caption=false]{subfig}
\usepackage{color}
\usepackage{booktabs}
\usepackage{tabularx}
\usepackage{multirow}
\usepackage[table]{xcolor}
\usepackage[normalem]{ulem}
\usepackage{dcolumn}
\usepackage{rotating}
\usepackage{hyperref}

\usepackage[latin1]{inputenc} 

\newcommand{\bi}{\begin{itemize}}
\newcommand{\ei}{\end{itemize}}

\def\beq{\begin{equation}}
\def\eeq{\end{equation}}

\newcommand{\bea}{\begin{eqnarray}}
\newcommand{\eea}{\end{eqnarray}}

\newcommand{\ldm}{\Delta m_{31}^2}
\newcommand{\sdm}{\Delta m_{21}^2}

\newcommand{\dcp}{\ensuremath{\delta}}

\newcommand{\ie}{{\it i.e.}}

\newcommand{\eet}{a_{e\tau}}

\newcommand{\eem}{a_{e\mu}}

\newcommand{\eetp}{\varphi_{e\tau}}

\newcommand{\eemp}{\varphi_{e\mu}}

\def\epsilon{\varepsilon}

\def\<{\langle}
\def\>{\rangle}

\def\lsim{\mathrel{\rlap{\lower4pt\hbox{\hskip1pt$\sim$}}
    \raise1pt\hbox{$<$}}}         
\def\gsim{\mathrel{\rlap{\lower4pt\hbox{\hskip1pt$\sim$}}
    \raise1pt\hbox{$>$}}}         

\begin{document}

\begin{titlepage}

\vspace*{-15mm}

\begin{flushright}
IP/BBSR/2019-9
\end{flushright}

\vspace*{0.8cm}

\begin{center}

{\bf\Large{Can Lorentz Invariance Violation affect the Sensitivity of \\
\vskip0.25cm
Deep Underground Neutrino Experiment?}} \\ [10mm]

{\bf Sanjib Kumar Agarwalla}$^{\, a,b,c,}$\footnote{E-mail: \texttt{sanjib@iopb.res.in} (ORCID: 0000-0002-9714-8866)},
{\bf Mehedi Masud}$^{\, a,d,}$\footnote{E-mail: \texttt{masud@ific.uv.es} (ORCID: 0000-0002-7014-3520)} \\
\vspace{8mm}
$^{a}$\,{\it Institute of Physics, Sachivalaya Marg, Sainik School Post, 
Bhubaneswar 751005, India} \\
\vspace{2mm}
$^{b}$\,{\it Homi Bhabha National Institute, Training School Complex, \\ 
Anushakti Nagar, Mumbai 400085, India} \\
\vspace{2mm}
$^{c}$\,{\it International Centre for Theoretical Physics, 
Strada Costiera 11, 34151 Trieste, Italy} \\
\vspace{2mm}
$^{d}$\,{\it Astroparticle and High Energy Physics Group, Instituto de F\'{i}sica Corpuscular (CSIC/Universitat de Val\`{e}ncia), 
Parc Cientific de Paterna. C/Catedratico Jos\'e Beltr\'an, 2 E-46980 Paterna (Val\`{e}ncia) - Spain}

\end{center}

\vspace{8mm}

\begin{abstract}
\vspace{4mm}
\noindent 
We examine the impact of Lorentz Invariance Violation (LIV) 
in measuring the octant of $\theta_{23}$ and CP phases in the context
of the Deep Underground Neutrino Experiment (DUNE). We consider the 
CPT-violating LIV parameters involving $e - \mu$ ($a_{e\mu}$) and
$e - \tau$ ($a_{e\tau}$) flavors, which induce an additional interference 
term in neutrino and antineutrino appearance probabilities. This new
interference term depends on both the standard CP phase $\delta$
and the new dynamical CP phase $\varphi_{e\mu}$/$\varphi_{e\tau}$,
giving rise to new degeneracies among ($\theta_{23}$, $\delta$, $\varphi$).
Taking one LIV parameter at-a-time and considering a small value of 
$|a_{e\mu}| = |a_{e\tau}| = 5 \times 10^{-24}$ 
GeV, we find that the octant discovery potential of DUNE
gets substantially deteriorated for unfavorable combinations of $\delta$
and $\varphi_{e\mu}$/$\varphi_{e\tau}$. The octant of $\theta_{23}$
can only be resolved at $3\sigma$ if the true value of 
$\sin^2\theta_{23} \lesssim 0.42$ or $\gtrsim 0.62$ for any choices 
of $\delta$ and  $\varphi$. Interestingly, we also observe that 
when both the LIV parameters $a_{e\mu}$ and $a_{e\tau}$
are present together, they cancel out the impact of each other
to a significant extent, allowing DUNE to largely regain its octant 
resolution capability. We also reconstruct the CP phases $\delta$ and 
$\varphi_{e\mu}$/$\varphi_{e\tau}$. The typical $1\sigma$
uncertainty on $\delta$ is $10^{\circ}$ to $15^{\circ}$ and 
the same on $\varphi_{e\mu}$/$\varphi_{e\tau}$ is $25^{\circ}$ 
to $30^{\circ}$ depending on the choices of their true values.

\end{abstract}
\end{titlepage}

\setcounter{footnote}{0}

\section{Introduction}
\label{sec:intro}

New opportunities have emerged on the neutrino oscillation frontier 
where almost all the available data can be nicely accommodated 
in a standard three-flavor oscillation framework~\cite{Tanabashi:2018oca},
except few intriguing anomalies uncovered by the short-baseline 
experiments (for recent reviews see~\cite{Boser:2019rta,Diaz:2019fwt}). 
There are six fundamental parameters in the three-neutrino (3$\nu$) mixing 
paradigm that govern the oscillation phenomena: 
a) three leptonic mixing angles ($\theta_{12}$, $\theta_{13}$, $\theta_{23}$),
b) one Dirac CP phase ($\delta$), and c) two distinct mass-squared 
splittings\footnote{$\Delta m^2_{21}$ ($\equiv m^2_2 - m^2_1$) governs
the oscillation in the solar sector and $\Delta m^2_{32}$ ($\equiv m^2_3 - m^2_2$)
is responsible for the oscillation in the atmospheric sector. Here, the neutrino 
mass eigenstate $m_3$ has the smallest electron component.}
($\Delta m^2_{21}$, $\Delta m^2_{32}$). After establishing the phenomena
of neutrino oscillation conclusively, neutrino physics has now entered into the
precision era with an aim to address the following three fundamental pressing 
issues at unprecedented confidence level.
\begin{itemize}

\item
Determining the value of charge-parity (CP) violating phase $\delta$ -- where
establishing a value differing from both zero and $\pi$ would symbolize the 
discovery of CP-violation (CPV) in the leptonic sector.

\item
Settling the pattern of neutrino masses. The present oscillation data 
cannot resolve whether $\Delta m^2_{31}$ ($\equiv m^2_3 - m^2_1$)
is positive or negative. It allows us to arrange the neutrino masses 
in two different fashions: $m_3 > m_2 > m_1$, called normal ordering (NO) 
where $\Delta m^2_{31}$ is positive and $m_2 > m_1 >  m_3$, 
known as inverted ordering (IO) where $\Delta m^2_{31}$ is negative. 

\item 
Precise measurement of the mixing angle $\theta_{23}$. If it turns out 
to be non-maximal ($\theta_{23} \neq \pi/4$), then we can have two
possibilities: $\theta_{23}$ can either lie in the lower octant (LO),  
\ie~$\theta_{23} < \pi/4$ or in the higher octant (HO), \ie~$\theta_{23} > \pi/4$. 

\end{itemize}

Presently running long-baseline neutrino oscillation experiments 
Tokai to Kamioka (T2K)~\cite{Abe:2013hdq} and NuMI Off-axis 
$\nu_e$ Appearance (NO${\nu}$A)~\cite{Ayres:2004js} have 
already started shading light on the above mentioned issues. 
Latest T2K results~\cite{Abe:2019vii} hint towards a HO value 
for $\sin^{2}\theta_{23}$ = $0.53^{+0.03}_{-0.04}$ for both NO
and IO. For the first time, T2K has been able to rule out a large 
range of values of $\delta$ around $\pi/2$ at $3\sigma$ 
C.L. irrespective of mass ordering. The CP conserving choices 
of $\delta$ (both $0$ and $\pi$) are also excluded at 95\% C.L. by 
the same data. The most recent measurements by the 
NO$\nu$A Collaboration~\cite{Acero:2019ksn} using both 
neutrinos and antineutrinos point towards NO, disfavoring 
IO at $1.9\sigma$ C.L. and shows a weak preference for 
$\theta_{23}$ in HO over LO at a C.L. of $1.6\sigma$.
The NO$\nu$A data excludes most of the choices near 
$\delta = \pi/2$ for IO at a C.L. $\geqslant 3\sigma$.
But these experiments still have a long way to go and 
hopefully, their results will be strengthened further with 
more statistics in near future. The global analyses of
world neutrino data~\cite{deSalas:2017kay, globalfit, Capozzi:2018ubv, Esteban:2018azc} 
also indicate towards NO at more than $3\sigma$ C.L. 
and a non-maximal $\theta_{23}$ around $2\sigma$ 
with a mild preference for HO. However, the value of 
the standard CP phase $\delta$ is still uncertain by 
a large extent.

The upcoming high-precision long-baseline neutrino 
oscillation experiments are expected to resolve these
crucial issues at high confidence level and to provide
a rigorous test of the three-flavor neutrino oscillation 
framework in the presence of Earth's matter
effect~\cite{Pascoli:2013wca,Agarwalla:2013hma,Agarwalla:2014fva}. 
These experiments include Deep Underground Neutrino 
Experiment (DUNE)~\cite{Acciarri:2015uup, Acciarri:2016ooe}, 
Tokai to Hyper-Kamiokande (T2HK)~\cite{Abe:2015zbg}, 
Tokai to Hyper-Kamiokande with a second detector in 
Korea (T2HKK)~\cite{Abe:2016ero}, and European 
Spallation Source $\nu$ Super Beam 
(ESS$\nu$SB)~\cite{Baussan:2013zcy,Agarwalla:2014tpa}.
These facilities are supposed to measure the mixing 
angles and mass-squared differences with a precision below 
{\it{few}} \% and therefore, these next generation neutrino experiments 
may be sensitive to various Beyond the Standard Model 
(BSM) scenarios~\cite{Arguelles:2019xgp}, which will 
complement the search for new physics at the ongoing 
LHC and future collider facilities. In this paper, we 
consider a specific BSM scenario of Lorentz Invariance 
Violation (LIV)~\cite{Kostelecky:2003cr, Diaz:2011ia} 
and analyze its impact on the measurements of 
$\theta_{23}$ octant and CPV at the DUNE facility.

The Standard Model (SM) is considered to be a low-energy 
effective gauge theory of a more fundamental framework that 
also unifies gravitational interactions along with Strong, 
Weak, and Electromagnetic interactions. The natural mass scale 
of that theory is governed by the Planck mass ($M_{P} \sim 10^{19}$ GeV). 
There exist studies that propose spontaneous LIV and CPT 
violations\footnote{In a seminal paper by O. W. Greenberg, 
it was shown that CPT violation implies violation of Lorentz 
Invariance~\cite{Greenberg:2002uu}.} in that more complete 
framework~\cite{Kostelecky:1988zi, Kostelecky:1989jp, Kostelecky:1991ak, Kostelecky:1994rn, Kostelecky:1995qk}. 
In the observable low-energy limit, this spontaneous violation 
of CPT/Lorentz symmetry can give rise to a minimal extension 
of the standard model through small perturbative terms 
suppressed by $M_{P}$. In the present work, we consider 
this minimal extension of the SM (as developed 
in~\cite{Colladay:1996iz, Colladay:1998fq, Kostelecky:2003cr, Diaz:2011ia, Kostelecky:2011gq}) 
which violates Lorentz invariance as well as CPT symmetry.

Neutrino experiments may determine the presence of Lorentz/CPT violation 
via the possible changes in neutrino oscillation probabilities, which can 
happen due to various reasons such as neutrino-antineutrino mixing, 
energy dependent effects on mass splittings, and time or direction 
dependent effects~\cite{Kostelecky:2003cr, Kostelecky:2004hg, Katori:2006mz, Diaz:2011ia}.
Several neutrino oscillation experiments such as
Liquid Scintillator Neutrino Detector (LSND)~\cite{Auerbach:2005tq}, 
Main Injector Neutrino Oscillation Search (MINOS)~\cite{Adamson:2008aa, Adamson:2010rn, Adamson:2012hp}, 
Mini Booster Neutrino Experiment (MiniBooNE)~\cite{AguilarArevalo:2011yi}, 
Double Chooz~\cite{Abe:2012gw}, Super-Kamiokande (SK)~\cite{Abe:2014wla}, 
IceCube~\cite{Abbasi:2010kx, Aartsen:2017ibm}, and T2K~\cite{Abe:2017eot}
have searched for these LIV/CPT-violating effects in their datasets and have
placed competitive constraints on these LIV/CPT-violating parameters.
Besides the above mentioned studies by the official Collaborations, 
there are also several other independent attempts on constraining 
LIV/CPT-violating parameters in the context of long-baseline accelerator 
neutrinos~\cite{Dighe:2008bu, Barenboim:2009ts, Rebel:2013vc, deGouvea:2017yvn, Barenboim:2017ewj, Barenboim:2018ctx, Majhi:2019tfi}, 
short-baseline reactor antineutrinos~\cite{Giunti:2010zs}, atmospheric neutrinos~\cite{Datta:2003dg, Chatterjee:2014oda, Koranga:2014dua}, 
solar neutrinos~\cite{Diaz:2016fqd}, and high-energy astrophysical neutrinos~\cite{Hooper:2005jp, Tomar:2015fha, Liao:2017yuy}. 
Hadron colliders such as LHC can also provide unique opportunity 
to test LIV/CPT-violating effects at high energy~\cite{Carle:2019ouy, Chanon:2019igm}.
A comprehensive list of the constraints on all the relevant 
LIV/CPT-violating parameters is available in Ref.~\cite{Kostelecky:2008ts}. 
In a recent work, the authors of~\cite{Barenboim:2018ctx} have performed 
a detailed analysis to put stronger bounds on the most relevant CPT-violating 
LIV parameters by simulating the upcoming DUNE experiment.   
Using these more tightly constrained CPT-violating LIV parameters, 
we study here for the first time, the octant sensitivity of DUNE 
in presence of the LIV parameters $(|\eem|, \eemp)$ and $(|\eet|, \eetp)$. 
Also, we have explored the capability of DUNE to reconstruct 
the true values of the standard Dirac CP phase $\delta$ 
and the LIV phases $\eemp$ and $\eetp$. Our analyses include 
the presence of both of these sets of LIV parameters individually 
as well as collectively. For recent status of searches of various 
BSM physics other than LIV using neutrino experiments, 
see~\cite{Farzan:2017xzy,Esteban:2019lfo,Boser:2019rta,Diaz:2019fwt,Dentler:2018sju,Escrihuela:2015wra,Blennow:2016jkn,Berryman:2016szd,Ballett:2019xoj} and the references therein.

This paper is organized as follows. 
Section~\ref{sec:theory} gives a short overview of the 
theoretical background pertaining to LIV scenario
and provides an analytical discussion on how
LIV parameters alter the neutrino and antineutrino 
appearance probability expressions by introducing 
an additional interference term, which depends on 
both the standard CP phase $\delta$ and 
the new dynamical CP phase 
$\varphi_{e\mu}$/$\varphi_{e\tau}$, giving rise to 
new degeneracies among $\theta_{23}$, $\delta$, 
and $\varphi$. In the same section, we also derive 
approximate analytical expressions to show how 
these new degeneracies affect the measurement 
of octant of $\theta_{23}$. Section~\ref{sec:simulation} 
discusses the important detector properties and 
the $\Delta \chi^{2}$ analysis procedure. 
In Sec.~\ref{sec:prob_event}, we show how various 
LIV parameters affect the exact numerical transition 
probability $P_{\mu e}$. In the same section, 
we also give bi-event plots to depict how much 
variation one can expect in the neutrino and 
antineutrino appearance event rates due to various 
LIV parameters. We present our main results 
concerning the octant discovery potential and 
the capability of reconstruction of the CP phases 
in Sec.~\ref{sec:results}. Finally, in 
Sec.~\ref{sec:summary}, we summarize our results 
and conclude. In appendix~\ref{sec:LIV-in-data-not-in-fit}, 
we explore the octant discovery potential of DUNE 
assuming the presence of LIV in data, but not in 
fit (theory).

\section{LIV formalism}
\label{sec:theory}
Lorentz Invariance violating neutrinos 
and antineutrinos are effectively described 
by the Lagrangian density~\cite{Kostelecky:2003cr, Kostelecky:2011gq},
\begin{align}\label{eq:lag}
\mathcal{L} = \frac{1}{2}\bar{\psi}({i\slashed{\partial}} - M + \hat{\mathcal{Q}})\psi + h.c.,
\end{align}
where, $\hat{\mathcal{Q}}$ is a generic 
Lorentz Invariance violating operator and 
the spinor $\psi$ describes the neutrino field. 
The first term on the right hand side (RHS) of 
Eq.~\ref{eq:lag} is the usual kinetic term, the second part 
involves the mass term with the mass matrix M 
and the 3rd term gives rise to the 
Lorentz Invariance violating effect,
which is small and perturbative in nature, 
possibly arising from Planck-suppressed effects. 
Considering only the renormalizable Dirac 
couplings of the theory, we can start from 
the Lorentz Invariance violating Lagrangian~\cite{Kostelecky:2011gq},
\begin{align}\label{eq:lag_cpt}
\mathcal{L}_{\text{LIV}} = -\frac{1}{2}\left[ a^{\mu}_{\alpha\beta}\bar{\psi}_{\alpha}\gamma_{\mu}\psi_{\beta} + b^{\mu}_{\alpha\beta}\bar{\psi}_{\alpha}\gamma_{5}\gamma_{\mu}\psi_{\beta} 
- i  c_{\alpha\beta}^{\mu\nu}   \bar{\psi}_{\alpha}\gamma_{\mu}\partial_\nu\psi_{\beta}
- i d_{\alpha\beta}^{\mu\nu}   \bar{\psi}_{\alpha}\gamma_5\gamma_{\mu}\partial_\nu\psi_{\beta}
\right] + h.c.
\end{align}
The observable effect on the left handed 
neutrinos is controlled by the combinations
\begin{align}
(a_{L})^{\mu}_{\alpha\beta} = (a + b)^{\mu}_{\alpha\beta} \, , \quad \quad  (c_{L})^{\mu\nu}_{\alpha\beta} = (c + d)^{\mu\nu}_{\alpha\beta}  \, ,
\end{align}
which are constant hermitian matrices in the flavor 
space that can modify the standard vacuum Hamiltonian. 
The first combination is relevant for CPT-violating neutrinos, 
whereas the second combination is only relevant for 
CPT-even Lorentz-violating neutrinos. In this work, we will 
focus on the isotropic component of the Lorentz-violating 
terms and therefore, we will fix the ($\mu$,$\nu$) indices 
to zero (0). To simplify our notation, from now on, we 
will denote\footnote{These components are defined in 
the Sun-centered celestial equatorial frame~\cite{Kostelecky:2003cr}.}
the parameter $(a_{L})^{0}_{\alpha\beta}$ as $a_{\alpha\beta}$ 
and $(c_{L})^{00}_{\alpha\beta}$ as $c_{\alpha\beta}$.

Explicitly, one can write the Lorentz-violating contribution 
to the full oscillation Hamiltonian
\begin{align}\label{eq:h_cpt}
H = H_{\text{vac}} + H_{\text{mat}} + H_{\text{LIV}},
\end{align}
such that,
\begin{align}\label{eq:si_part}
H_{\text{vac}} = \frac{1}{2E}U
\left(\begin{array}{ccc} m_{1}^{2} & 0 & 0 \\ 0 & m_{2}^{2} & 0 \\ 0 & 0 & m_{3}^{2} \\\end{array}\right)U^{\dagger}; 
\qquad H_{\text{mat}} = \sqrt{2}G_{F}N_{e}\left(\begin{array}{ccc} 1 && \\ &0& \\ &&0\\ \end{array}\right);
\end{align}
\begin{align}\label{eq:cpt_part}
H_{\text{LIV}} = 
\left(
\begin{array}{ccc}
a_{ee} & a_{e\mu} & a_{e\tau} \\
a^*_{e\mu} & a_{\mu\mu} & a_{\mu\tau} \\
a^*_{e\tau} & a^*_{\mu\tau} & a_{\tau\tau}
\end{array}
\right)
-\frac{4}{3}
E \left(
\begin{array}{ccc}
c_{ee} & c_{e\mu} & c_{e\tau} \\
c^*_{e\mu} & c_{\mu\mu} & c_{\mu\tau} \\
c^*_{e\tau} & c^*_{\mu\tau} & c_{\tau\tau}
\end{array}
\right),
\end{align}
where, $U$ is the neutrino mixing matrix, 
$m_{i}$'s are the neutrino mass eigenstates, 
$G_{F}$ is the Fermi coupling constant, and 
$N_{e}$ is the electron density along the 
neutrino trajectory. The $a_{\alpha\beta}$'s 
and $c_{\alpha\beta}$'s are the LIV parameters. 
In Eq.\ \ref{eq:cpt_part}, the factor $-4/3$ in front 
of the second term arises from the non-observability 
of the Minkowski trace of  $c_L$, which forces 
the components $xx$, $yy$, and $zz$ to be related 
to the 00 component~\cite{Kostelecky:2003cr}. 
In this work, we consider the presence of 
Lorentz-violating effects only due to the first 
type of terms\footnote{Therefore, we can argue
that the LIV scenario which we analyze in the
present work is also CPT-violating in nature.}
in Eq.~\ref{eq:cpt_part}. Updated constraints 
on $a_{\alpha\beta}$'s, mainly from Super-Kamiokande, 
can be found in Refs.~\cite{Abe:2014wla,Kostelecky:2008ts}.
Note that, after considering only the CPT-violating 
LIV terms ($a_{\alpha\beta}$'s), the LIV effect looks 
similar to the effect of neutral current (NC) non-standard 
interaction (NSI) during neutrino propagation, which can 
be described in the following fashion
\begin{equation}\label{eq:h_nsi}
H^{'} = H_{\text{vac}} + H_{\text{mat}} + H_{\text{NSI}}\, ,
\end{equation}
where the NSI term is parameterized as
\begin{equation}\label{eq:nsi_part}
H_{\text{NSI}} = 
\sqrt{2}G_{F} N_{e}\left(
 \begin{array}{ccc}
\epsilon^m_{ee} & \epsilon^{m}_{e\mu} & \epsilon^{m}_{e\tau} \\
\epsilon^{m}_{\mu e} & \epsilon^{m}_{\mu\mu} & \epsilon^{m}_{\mu\tau} \\
\epsilon^{m}_{\tau e} & \epsilon^{m}_{\tau\mu} & \epsilon^{m}_{\tau\tau}
\end{array}
\right) 
\, .
\end{equation}
Here, $N_e$ corresponds to the electron number density 
along the neutrino trajectory and the parameters 
$\epsilon^m_{\alpha \beta}$ denote the strength 
of the NSI. One thus finds a correlation between 
the NSI and LIV scenario through the following 
relation~\cite{Diaz:2015dxa},
\begin{equation}\label{eq:nsi_cpt}
\epsilon^{m}_{\alpha\beta}\  \equiv \frac{a_{\alpha\beta}}{\sqrt{2}G_{F} N_{e}}.
\end{equation}
However, there are important differences between 
these two scenarios~\cite{Diaz:2015dxa, Barenboim:2018lpo}. 
NSI during neutrino propagation is basically an exotic matter 
effect and hence, plays no role in vacuum, whereas the type 
of LIV considered here is an intrinsic effect, present even in 
vacuum. Nevertheless, the equivalence in Eq.~\ref{eq:nsi_cpt} 
allows the study of the LIV parameters in long-baseline 
experiments following an approach, which is quite similar 
to the treatment of NSI in neutrino propagation.

In this paper, we only consider the LIV parameters 
$\eem$ ($\equiv|\eem|e^{i\eemp}$) and $\eet$ ($\equiv|\eet|e^{i\eetp}$) 
since these parameters influence the most $\nu_{\mu} \to \nu_{e}$ 
appearance channel, which drives the CPV and octant sensitivity 
in a typical long-baseline experiment such as DUNE.
The probability expression for $\nu_{\mu} \to \nu_{e}$ 
oscillation channel in presence of the LIV parameters 
$\eem$ and $\eet$ can be written as (following the similar 
expressions in presence of the NSI parameters 
$\varepsilon_{e\mu}$ and $\varepsilon_{e\tau}$
in Refs.~\cite{Kikuchi:2008vq, Agarwalla:2016fkh,Masud:2018pig}):
\begin{align}\label{eq:pme_liv}
P_{\mu e} \simeq P_{\mu e}(\text{SI}) + P_{\mu e}(\eem) + P_{\mu e}(\eet),
\end{align}
where, the three terms on the RHS are described below. 
The first term originating from the standard interaction 
(SI) of neutrinos with the Earth's matter is given by
\begin{align}\label{eq:p_si}
P_{\mu e}(\text{SI}) 
&\simeq X + Y\cos(\delta + \Delta),
\end{align}
where,
\begin{align}\label{eq:si_coeff}
&X = 4s_{13}^{2}c_{13}^{2} s_{23}^{2} \frac{\sin^{2}\big[(1-\hat{A})\Delta \big]}{(1-\hat{A})^{2}}; \qquad
Y = 8\alpha s_{12} c_{12} s_{23} c_{23} s_{13}c_{13} \frac{\sin \hat{A}\Delta}{\hat{A}}  \frac{\sin \big[(1-\hat{A})\Delta\big]}{1-\hat{A}}, \nonumber \\
&\hat{A} = \frac{2\sqrt{2}G_{F}N_{e}E}{\ldm}; \qquad  \Delta = \frac{\ldm L}{4E}; \qquad s_{ij} = \sin\theta_{ij}; \qquad c_{ij} = \cos\theta_{ij}; \qquad \alpha = \frac{\sdm}{\ldm}.
\end{align}
In writing the expression for $P_{\mu e}(\text{SI})$ 
in Eq.~\ref{eq:p_si}, we neglect the {\it{solar}} term 
$\alpha^{2} \sin^{2} 2\theta_{12} c_{23}^{2} \frac{\sin^{2}\hat{A}\Delta}{\hat{A}^{2}}$.
This is due to the fact that by considering the values 
of the oscillation parameters as $\theta_{12} = 34.5^{\circ}, 
\theta_{13} = 8.45^{\circ}, \theta_{23} = 47.7^{\circ}, 
\sdm = 7.5 \times 10^{-5} \text{ eV}^{2}, 
\ldm = 2.5 \times 10^{-3} \text{ eV}^{2}$ 
(which are in agreement with~\cite{deSalas:2017kay,globalfit,Capozzi:2018ubv,Esteban:2018azc}), 
we find that the {\it{solar}} term, being proportional to $\alpha^{2}$, 
is roughly suppressed by 3 to 4 orders of magnitude 
as compared to the other two terms as shown 
in Eq.~\ref{eq:p_si}.

To describe the second and the third terms of the RHS 
of Eq.~\ref{eq:pme_liv}, describing the effect of LIV due 
to the presence of $\eem$ and $\eet$ respectively, 
we take the similar approach as followed in the context 
of NC NSI in~\cite{Kikuchi:2008vq, Agarwalla:2016fkh,Masud:2018pig} 
with the NSI parameter $\varepsilon_{\alpha\beta}$ replaced 
appropriately (as in Eq.~\ref{eq:nsi_cpt}). 
Thus, the LIV terms in Eq.\ \ref{eq:pme_liv} can be written 
in the following compact form:
\begin{align}\label{eq:p_liv}
P_{\mu e}(a_{e\beta}) 
\simeq \frac{4 |a_{e\beta}| \hat{A} \Delta s_{13} \sin2\theta_{23} \sin\Delta}{\sqrt{2}G_{F}N_{e}}\big[Z_{e\beta}\sin(\delta + \varphi_{e\beta}) +
W_{e\beta}\cos(\delta + \varphi_{e\beta})\big], \quad(\beta = \mu, \tau)
\end{align}
where,
\begin{align}\label{eq:liv_coeff}
  &Z_{e\beta} = 
  \begin{cases}
      - c_{23}  \sin \Delta, & \text{if}\ \beta=\mu. \\
       s_{23}  \sin \Delta, & \text{if}\ \beta=\tau.
    \end{cases} \nonumber \\
 &W_{e\beta} = 
  \begin{cases}
       c_{23}  \big(\frac{s_{23}^{2}\sin \Delta}{c_{23}^{2}\Delta} + \cos \Delta \big), & \text{if}\ \beta=\mu. \\
       s_{23}  \big(\frac{\sin \Delta}{\Delta} - \cos \Delta \big), & \text{if}\ \beta=\tau.
    \end{cases} 
\end{align}
We note that the factor 
$\frac{\hat{A}\Delta}{\sqrt{2}G_{F}N_{e}}$ 
in Eq.~\ref{eq:p_liv} becomes $L/2$ 
(see Eq.~\ref{eq:si_coeff}.), thereby 
making the LIV effect considered here 
explicitly matter independent.

Following~\cite{Agarwalla:2016fkh}, 
now we explain the issue of octant
sensitivity of $\theta_{23}$ by expressing 
the atmospheric mixing angle as:
\begin{align}\label{eq:th23_pert}
\theta_{23} = \frac{\pi}{4} \pm \eta,
\end{align}
such that the positive angle $\eta$ 
quantifies the deviation from the 
maximal mixing. The positive (negative) 
sign corresponds to HO (LO). The current 
three-flavor global analyses~\cite{deSalas:2017kay, globalfit, Capozzi:2018ubv, Esteban:2018azc} 
indicate that $\theta_{23}$ cannot deviate from 
$45^{\circ}$ by more than $\sim 6^{\circ}$, \ie, 
$\sin^2\theta_{23}$ must be in the range 
$[0.4, 0.6]$. Therefore, one has 
$\eta \lesssim 0.1$, and 
we can use the expansions, 
\begin{align}
s_{23}^{2} \simeq \frac{1}{2} \pm \eta;
\qquad  c_{23}^{2} \simeq \frac{1}{2} \mp \eta; 
\qquad \sin2\theta_{23} \simeq 1.
\end{align}
An experiment is sensitive to the 
octant of $\theta_{23}$ if at the probability 
level the following difference between 
the {\it{true octant}} (tr) and {\it{test octant}} (ts) 
is nonzero at a detectable level.
\begin{equation}
\Delta P = P_{\mu e}^{\text{tr}} (\theta_{23}^{\text{tr}}, \delta^{\text{tr}}, \varphi^{\text{tr}})
- P_{\mu e}^{\text{ts}} (\theta_{23}^{\text{ts}}, \delta^{\text{ts}}, \varphi^{\text{ts}})\,.
\end{equation}
Since $P_{\mu e}$ consists of the three terms 
in Eq.~\ref{eq:pme_liv}, we can write, 
\begin{equation}\label{eq:D_P}
\Delta P \simeq \Delta P_{\mu e}(\text{SI}) + \Delta P_{\mu e}(\eem) + \Delta P_{\mu e}(\eet)\,.
\end{equation}
To analyse the three terms on the RHS 
of Eq.~\ref{eq:D_P}, we take the case of 
HO (LO) as {\it{true}} ({\it{test}}) octant 
as an example. Then for the SI term,
\begin{equation}\label{eq:D_P_si}
\Delta P_{\mu e}(\text{SI}) = \Delta X + \Delta Y \big[
\cos (\delta^{\text{HO}} + \Delta) - \cos (\delta^{\text{LO}} + \Delta)
\big],
\end{equation}
where,
\begin{equation}\label{eq:D_P_si_coeff}
\Delta X \simeq 8 \eta s_{13}^{2} c_{13}^{2} \frac{\sin^{2}\big[(1-\hat{A})\Delta\big]}{(1-\hat{A})^{2}}; \qquad
\Delta Y \simeq 4\alpha s_{12} c_{12}  s_{13}c_{13} \frac{\sin \hat{A}\Delta}{\hat{A}}  \frac{\sin\big[(1-\hat{A})\Delta\big]}{1-\hat{A}}\,.
\end{equation}
The LIV contribution to Eq.~\ref{eq:D_P} 
can be written as,
\begin{align}\label{eq:D_P_liv}
\Delta P_{\mu e}(a_{e\beta}) \simeq \frac{4 |a_{e\beta}| \hat{A} \Delta s_{13} \sin2\theta_{23} \sin\Delta}{\sqrt{2}G_{F}N_{e}}\bigg[
\Delta Z_{e\beta}&\bigg\{
\sin(\delta^{\text{HO}} + \varphi_{e\beta}^{\text{HO}})
- \sin(\delta^{\text{LO}} + \varphi_{e\beta}^{\text{LO}})
\bigg\}  \nonumber \\
{} + \Delta W_{e\beta}&\bigg\{
\cos(\delta^{\text{HO}} + \varphi_{e\beta}^{\text{HO}}) 
- \cos(\delta^{\text{LO}} + \varphi_{e\beta}^{\text{LO}})
\bigg\}
\bigg],
\end{align}
where, 
\begin{align}\label{eq:D_P_liv_coeff}
&\Delta Z_{e\beta} \simeq \mp \frac{1}{\sqrt{2}}\sin \Delta, \quad \text{[where the $- (+)$ sign is for $\beta = \mu (\tau)$],} \nonumber \\
&\Delta W_{e\beta} \simeq 
\begin{cases}
\frac{1}{\sqrt{2}}\big[ \frac{\sin \Delta}{\Delta} + \cos \Delta \big], & \text{if}\ \beta=\mu, \\
\frac{1}{\sqrt{2}}\big[ \frac{\sin \Delta}{\Delta} - \cos \Delta \big], & \text{if}\ \beta=\tau.
\end{cases}
\end{align}

In DUNE, neutrinos and antineutrinos travel the distance 
of $L = 1300$ km and for this baseline, the line-averaged
constant Earth matter density turns out to be $\rho = 2.95$ 
g/cm$^3$~\cite{Roe:2017zdw}. We also assume that Earth's 
matter is electrically neutral and isoscalar for which we have
$N_e = N_p = N_n$, where $N_p$, $N_n$ are the proton 
and neutron densities respectively. Under this assumption, 
the relative number density $Y_e$ 
($\equiv {N_e \over N_p + N_n}$) comes out to be 0.5.
Also note that for DUNE baseline, the first oscillation 
maximum ($\Delta \simeq \pi/2$) occurs at $E \approx 2.5$ 
GeV assuming $\ldm = 2.5 \times 10^{-3} \text{ eV}^{2}$. 
With these benchmark choices of parameters,
we obtain the following approximate numerical values:
\begin{align}\label{eq:coeff_num}
&\Delta \simeq \pi/2, \\ 
&\sqrt{2}G_{F}N_{e} \simeq [7.6 \times Y_e \times 10^{-14} \times \rho~\text{(g/cm}^3)]~\text{eV} \simeq 1.12 \times 10^{-13} \text{eV}, \nonumber \\
&\hat{A} = \frac{2\sqrt{2}G_{F}N_{e}E}{\ldm} \simeq  0.23, \nonumber \\
&\frac{\sin(1-\hat{A})\Delta}{1-\hat{A}} \simeq 1.21, \nonumber \\
& \frac{\sin \hat{A}\Delta}{\hat{A}} \simeq 1.54. \nonumber
\end{align}
Now, to have an idea about the magnitude 
of the coefficients in Eqs.\ \ref{eq:D_P_si} 
and \ref{eq:D_P_liv}, we use the values of 
the oscillation parameters mentioned before 
and also Eqs.~\ref{eq:D_P_si_coeff}, 
\ref{eq:D_P_liv_coeff}, \ref{eq:coeff_num},
and obtain the following at the 1st oscillation
maxima: 
\begin{align}\label{eq:D_P_si_num}
\Delta P_{\mu e}(\text{SI}) \simeq  \frac{\eta}{0.05}1.26 \times 10^{-2} + 
1.5 \times 10^{-2}\big[ \cos (\delta^{\text{HO}} + \Delta) - \cos (\delta^{\text{LO}} + \Delta) \big] \,,
\end{align}
\begin{align}\label{eq:D_P_liv_num}
\Delta P_{\mu e}(a_{e\beta}) \simeq \bigg[\frac{|a_{e\beta}|\text{GeV}^{-1}}{5 \times 10^{-24}}\bigg]\bigg[
& \mp 0.67 \times 10^{-2} \bigg\{
\sin(\delta^{\text{HO}} + \varphi_{e\beta}^{\text{HO}}) 
- \sin(\delta^{\text{LO}} + \varphi_{e\beta}^{\text{LO}})
\bigg\} \nonumber \\
&+ 0.42 \times 10^{-2}\bigg\{
\cos(\delta^{\text{HO}} + \varphi_{e\beta}^{\text{HO}}) 
- \cos(\delta^{\text{LO}} + \varphi_{e\beta}^{\text{LO}})
\bigg\}
\bigg],
\end{align}
where, $-(+)$ sign is for $\beta = \mu (\tau)$.
It is clear from Eqs.~\ref{eq:D_P_si_num} and 
\ref{eq:D_P_liv_num} that for $|\eem| (|\eet|)
\gtrsim 10^{-24}$ GeV, $\Delta P_{\mu e}(\eem)$ 
($\Delta P_{\mu e}(\eet)$) becomes comparable 
to the standard interference term in 
$\Delta P_{\mu e}(\text{SI})$. Moreover, 
$\Delta P_{\mu e}(\eem)$ and 
$\Delta P_{\mu e}(\eet)$ depend not only on the 
standard CP phase $\delta$, but also on the new 
dynamical CP phase $\varphi_{e\mu}$/$\varphi_{e\tau}$ 
related to the LIV. Due to this extra degree of freedom
in $\Delta P_{\mu e}(\eem)$/$\Delta P_{\mu e}(\eet)$,
the octant sensitivity can potentially become worse
for unfavorable combinations of $\delta$
and $\varphi_{e\mu}$/$\varphi_{e\tau}$.
In addition, we note that the first terms in
$\Delta P_{\mu e}(\eem)$ and $\Delta P_{\mu e}(\eet)$
appear with the opposite sign. It suggests that
when both the LIV parameters $\eem$ and $\eet$ 
are present together, their effect may get cancelled 
to a large extent, and the chances of measuring 
octant in DUNE remain intact.

\section{Simulation details}
\label{sec:simulation}

The proposed Deep Underground Neutrino Experiment (DUNE)
is a world-class facility which is going to unravel some fundamental
issues in neutrino sector, namely, the measurement of leptonic 
CP-violation, the determination of the neutrino mass ordering, 
and the precision measurement of the neutrino mixing 
parameters~\cite{Acciarri:2015uup,Acciarri:2016ooe,Acciarri:2016crz,Abi:2018dnh}. 
In order to simulate DUNE, we use the GLoBES package~\cite{Huber:2004ka,Huber:2007ji} 
with the most recent DUNE configuration files provided by the collaboration~\cite{Alion:2016uaj}. 
To analyze the Lorentz-violating scenario, we perform our simulation of the DUNE 
experiment using the GLoBES-extension \textit{snu.c} as described 
in Refs.~\cite{Kopp:2006wp,Kopp:2007ne}. This extension was 
originally introduced in GLoBES software to study non-standard neutrino 
interactions and sterile neutrinos in the context of long-baseline experiments. 
For the present analysis, we modify the definition of the neutrino oscillation 
probability function inside \textit{snu.c} by implementing the Lorentz-violating 
Hamiltonian as given in Eq.~\ref{eq:cpt_part}. We assume a total run-time 
of 7 years with 3.5 years in the neutrino mode and the remaining 3.5 years 
in the antineutrino mode with an on-axis 40 kiloton liquid argon far detector 
(FD) housed at the Homestake Mine in South Dakota over a baseline 
of 1300 km. The optimized neutrino beam is obtained from a G4LBNF 
simulation~\cite{Agostinelli:2002hh,Allison:2006ve} of the LBNF 
beam line using NuMI-style focusing. The neutrino beam
is generated using 80 GeV proton beam having a beam 
power of 1.07 MW, which can deliver $1.47 \times 10^{21}$ 
protons on target per calendar year. It corresponds to a total 
exposure of 300 kt$\cdot$MW$\cdot$yrs.   

To simulate the DUNE event spectra, we consider the reconstructed 
neutrino and antineutrino energy range of 0 to 20 GeV for both
appearance and disappearance channels. While preparing our 
sensitivity plots, we perform a full spectral analysis with total 
$71$ bins in the entire energy range having non-uniform bin
widths. We have total $64$ bins each having a width of 
$0.125$ GeV in the energy range of 0 to 8 GeV and $7$ bins
with variable widths beyond $8$ GeV~\cite{Alion:2016uaj}.
While estimating the signal and background event rates 
in the appearance and disappearance modes, we properly
take into account the ``wrong-sign" components, which are
present in the beam. We do so for both $\nu_e$/$\bar\nu_e$
and $\nu_{\mu}$/$\bar\nu_{\mu}$ candidate events.
We calculate the full three-flavour neutrino oscillation 
probabilities in matter considering the line-averaged 
constant Earth matter density of 2.95 g/cm$^3$ following
the standard Preliminary Reference Earth Model 
(PREM)~\cite{Dziewonski:1981xy}.

The main sources of backgrounds for the appearance 
events in neutrino and antineutrino modes are the 
intrinsic $\nu_e$/$\bar\nu_e$ contamination in the 
beam, the $\mu^{-}/\mu^{+}$ events which are 
misidentified as $e^{-}/e^{+}$ events, backgrounds 
arising from $\nu_{\tau}$/$\bar\nu_{\tau}$ appearance,
and the NC events. For the disappearance events in 
neutrino and antineutrino modes, the main backgrounds
stem from the NC events and $\nu_{\tau}$/$\bar\nu_{\tau}$ 
appearance. We incorporate the systematic uncertainties
following Ref.~\cite{Alion:2016uaj}. We consider an 
independent normalization uncertainty of 2\% on both 
$\nu_e$ and $\bar\nu_e$ signal modes, while the 
$\nu_{\mu}$ and $\bar\nu_{\mu}$ signal modes have
uncorrelated normalization errors of 5\%. As far as the
normalization uncertainties on various backgrounds 
are concerned, they vary in the range of 5\% to 20\%
with possible correlations among various sources of
backgrounds.

To obtain the sensitivity results, we numerically calculate 
the $\Delta \chi^{2}$ between the {\it{true}} and {\it{test}} 
event spectra using GLoBES. Unless mentioned otherwise,
we consider three different benchmark values of
$|a_{e\beta}|$ = $10^{-24}$ GeV, $5 \times 10^{-24}$ GeV,
and $10^{-23}$ GeV (where $\beta$ can be $\mu$ or $\tau$)
while generating the {\it{true}} event spectra. While showing
our results in Sec.~\ref{sec:results}, we always marginalize
over the test $|a_{e\beta}|$ (where $\beta$ can be $\mu$ or $\tau$)
in the range of $5 \times 10^{-25}$ GeV to $5 \times 10^{-23}$ GeV
in the fit. For a true value of $\delta$ and $\varphi_{e\beta}$, the 
true number of events in the $i$-th energy bin
$N_{i}(\theta_{23}^{\textrm{true}}, \delta^{\textrm{true}}, 
\varphi_{e\beta}^{\textrm{true}})$ is estimated
by assuming a true octant scenario, which can be either 
true lower octant (\ie, $\theta_{23}^{\textrm{true}} < \pi/4$) 
or true higher octant (\ie, $\theta_{23}^{\textrm{true}} > \pi/4$). 
The fixed true and test values of the solar oscillation parameters
and 1-3 mixing angle are $\theta_{12} = 34.5^{\circ}$,
$\sdm = 7.5 \times 10^{-5} \text{ eV}^{2}$, and 
$\theta_{13} = 8.45^{\circ}$. As far as the atmospheric 
mass-squared difference is concerned, we generate the
data with a true value of $\ldm = 2.5 \times 10^{-3} \text{ eV}^{2}$
and we marginalize over test $\ldm$ in the fit in its
present 3$\sigma$ allowed range of $(2.41 - 2.6) \times 10^{-3}
\text{ eV}^{2}$. We assume NO both in data and 
theory\footnote{Since the sensitivity of DUNE to exclude 
the wrong mass ordering is very high~\cite{Abi:2018dnh}, 
we keep the ordering same in both {\it{true}} and {\it{test}} 
datasets while performing our simulation. We have checked 
that DUNE can discriminate between NO and IO at high
confidence level even in presence of the LIV parameters.
It becomes possible due to the crucial spectral information 
provided by the on-axis wide-band muon-(anti)neutrino 
beam in DUNE.}. The theoretical event spectra are generated 
assuming the opposite/wrong octant scenario, where 
$\theta_{23}^{\textrm{test}}$ is marginalized over 
all possible values in the wrong octant only. 
$\delta^{\textrm{test}}$ and $\varphi_{e\beta}^{\textrm{test}}$ 
are marginalized over the full parameter space of $[-\pi,\pi]$.
$|a_{e\beta}|^{\textrm{test}}$ is marginalized in the range 
of $5 \times 10^{-25}$ GeV to $5 \times 10^{-23}$ GeV as
mentioned above. The $\Delta \chi^{2}$ thus gives a quantitative 
idea about the capability of the experiment to distinguish 
the true octant scenario from the wrong 
octant\footnote{The $\Delta \chi^{2}$ is calculated 
using the method of pull~\cite{Huber:2002mx,Fogli:2002pt,GonzalezGarcia:2004wg,Gandhi:2007td}. 
Also, this $\Delta \chi^{2}$ is valid in the frequentist method 
of hypotheses testing~\cite{Fogli:2002pt, Qian:2012zn}.}.

\section{Transition probability and bi-event plots}
\label{sec:prob_event}

\begin{figure}[htb]
\centering
\includegraphics[width=\textwidth,height=0.8\textwidth]{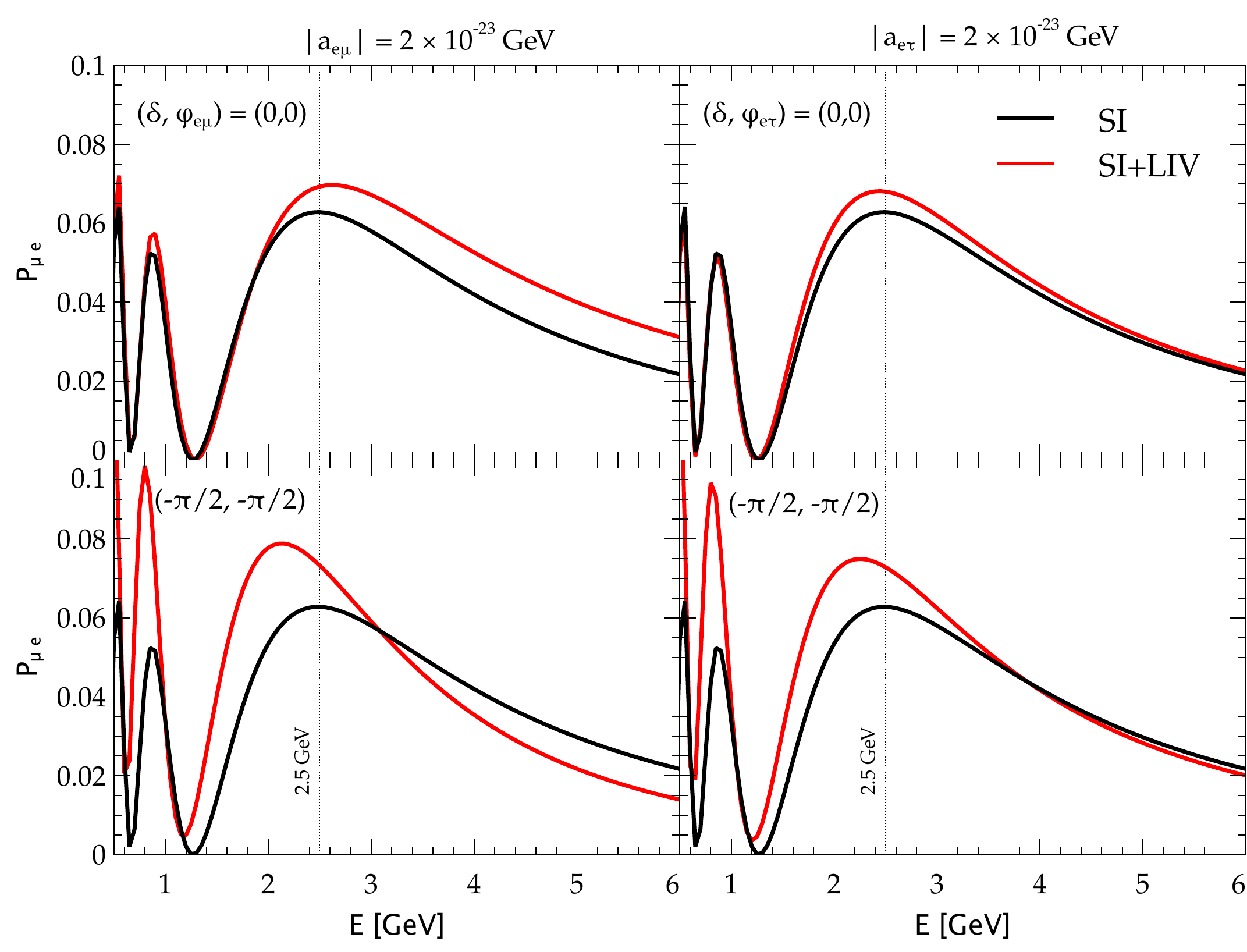}
\caption{\footnotesize{$\nu_{\mu} \to \nu_e$ 
transition probability as a function of neutrino
energy. In each panel, black curve shows the
probability considering only the standard interaction
(SI). The red curves depict how the LIV parameters
$a_{e\mu}$ (left panels) and $a_{e\tau}$ (right panels)
affect $P_{\mu e}$. The strength of the LIV parameters 
is assumed to be |$a_{e\mu}$| = |$a_{e\tau}$| = 
$2 \times 10^{-23}$ GeV (taken one at-a-time). 
In top (bottom) panels, we assume CP-conserving 
(CP-violating) values of the two relevant phases 
$\delta$ and $\varphi_{e\beta}$, whose values are
mentioned in each panel. Here, we assume NO 
and for the three-flavor oscillation parameters, 
we take the values $\theta_{12} = 34.5^{\circ},
\theta_{13} = 8.45^{\circ}, \theta_{23} = 47.7^{\circ}, 
\sdm = 7.5 \times 10^{-5} \text{ eV}^{2}$, and
$\ldm = 2.5 \times 10^{-3} \text{ eV}^{2}$.}}
\label{fig:prob}
\end{figure}

To demonstrate the impact of LIV, we have plotted 
$P_{\mu e}$ as a function of energy in Fig.~\ref{fig:prob} 
for both SI (black curves) and in presence of LIV (red curves). 
The left (right) panels assume the presence of the single 
LIV parameter $\eem$ ($\eet$). The top panels refer to
representative CP conserving values [0, 0] of the two relevant 
phases [$\delta, \varphi_{e\beta}$] as indicated in each panel. 
The bottom panels are for representative maximal CP-violating
choices [-$\pi/2$, -$\pi/2$] of the two CP phases 
[$\delta, \varphi_{e\beta}$] as mentioned in each panel.
For the purpose of illustration, here, we consider a 
relatively large strength of the LIV parameter 
($|\eem|$ or $|\eet|$ is taken to be $2 \times 10^{-23}$ GeV).
As far as the three-flavor oscillation parameters
are concerned, we consider the values 
$\theta_{12} = 34.5^{\circ}, \theta_{13} = 8.45^{\circ}, 
\theta_{23} = 47.7^{\circ}, \sdm = 7.5 \times 10^{-5} 
\text{ eV}^{2}, \ldm = 2.5 \times 10^{-3} 
\text{ eV}^{2}$, and assume NO.
Figure~\ref{fig:prob} clearly demonstrates that
the modifications in $\nu_\mu \to \nu_e$ transition 
probability due to the presence of LIV parameters
depend upon the values of the CP phases 
($\delta, \varphi_{{e\mu}/{e\tau}}$). The excellent
energy resolution in DUNE may enable us to study
the changes in the reconstructed event spectra due
to different choices of the CP phases 
($\delta, \varphi_{{e\mu}/{e\tau}}$), which in turn, 
may help us to reconstruct the values of these 
CP phases with reasonable accuracy. 

\begin{figure}[h!]
\centering
\includegraphics[scale=0.49]{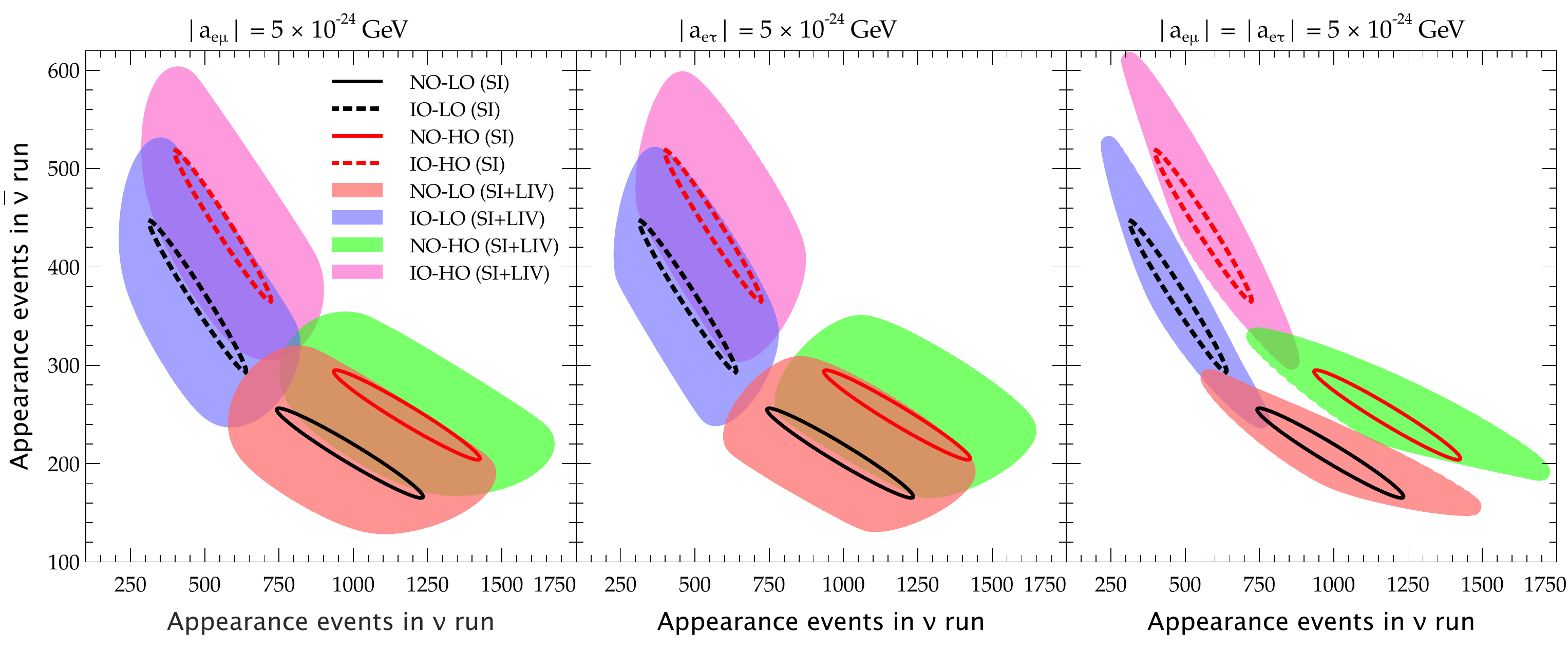}
\caption{\footnotesize{Bi-event plots for DUNE.
The standard interaction (SI) case is represented
by the solid/dashed ellipses, which are obtained
varying the standard CP phase $\delta$ in the 
range $[-\pi, \pi]$. The colored blobs denote the 
presence of LIV on top of the SI. These blobs are
generated varying the CP phases:
$\delta, \varphi_{e\mu}$ (left panel),
$\delta, \varphi_{e\tau}$ (middle panel),
$\delta, \varphi_{e\mu}$, and 
$\varphi_{e\tau}$ (right panel).
In all the cases, CP phases are allowed 
to vary in their entire ranges of [$-\pi, \pi$].
In all the panels, the strength of the LIV 
parameter is taken to be $5 \times 10^{-24}$
GeV. For both SI and SI+LIV, we consider 
four possible cases: two possible mass 
orderings (NO and IO) together with two possible 
octants (LO and HO), as shown in the legends.}}
\label{fig:bi_event}
\end{figure}

Now, we introduce the bi-event plots (see Fig.~\ref{fig:bi_event})
in which the x-axis (y-axis) denotes the total number of appearance
events in neutrino (antineutrino) mode. In all the panels, 
the solid/dashed ellipses depict the SI case, while the colored
blobs portray the SI+LIV scheme. The ellipses are obtained 
by varying the standard Dirac CP phase $\delta$ in the range
$[-\pi, \pi]$. In case of SI+LIV, there are more than one CP
phase and due to the simultaneous variation of these phases 
in their allowed ranges, we have a convolution of an infinite 
ensemble of ellipses with different orientations, which give rise 
to the colored blobs, In left (middle) panel, we obtain the blobs
by varying the CP phases $\delta, \varphi_{e\mu}$ 
($\delta, \varphi_{e\tau}$). In right panel, we vary three CP 
phases $\delta, \varphi_{e\mu}$, and $\varphi_{e\tau}$ 
at the same time in their allowed ranges of [$-\pi, \pi$].
In all the panels, the strength of the LIV parameter 
is taken to be $5 \times 10^{-24}$ GeV, which is 
consistent with~\cite{Barenboim:2018ctx}.
For both SI and SI+LIV, we study four possible 
cases: two possible mass orderings (NO and IO) 
together with two possible octants (LO and HO), 
as mentioned in the figure legends. The black
solid (dashed) ellipse correspond to the NO-LO
(IO-LO) case, while the red solid (dashed) ellipse
represents the NO-HO (IO-HO) scenario. In case
of NO, we take the values of the oscillation 
parameters as $\theta_{12} = 34.5^{\circ}, 
\sdm = 7.5 \times 10^{-5} \text{ eV}^{2},
\theta_{13} = 8.45^{\circ}, \ldm = 2.5 \times 10^{-3} 
\text{eV}^{2}$, and $\theta_{23} = 42.3^{\circ}$ 
($47.7^{\circ}$) for LO (HO) case~\cite{deSalas:2017kay,globalfit}. 
For the IO case, the values of the solar oscillation parameters 
($\theta_{12}$ and $\sdm$) remain the same and for the remaining 
oscillation parameters, we consider $\theta_{13} = 8.53^{\circ}, 
\ldm = -2.42 \times 10^{-3} \text{eV}^{2}$, and $\theta_{23} = 42.1^{\circ}$ 
($47.9^{\circ}$) for LO (HO) scenario~\cite{deSalas:2017kay,globalfit}.
For the SI case, there is a clear separation between 
the black (LO) and red (HO) ellipses for both NO and IO.
Once we introduce the LIV parameters $\eem$ (left panel) 
and $\eet$ (middle panel) one at-a-time, the LO and HO
blobs show significant overlap among each other for both 
possible mass orderings. It suggest that in presence 
of a single LIV parameter, the $\theta_{23}$ octant 
separation capability of DUNE may get deteriorated
significantly, which we confirm with the help of octant 
sensitivity plots in the next section. Interestingly, when 
both the LIV parameters $\eem$ and $\eet$ are present 
together in the scenario (see right panel of 
Fig.~\ref{fig:bi_event}), the amount of overlap between 
the LO and HO blobs gets reduced considerably 
as compared the single LIV parameter case.
This feature is consistent with our previous discussion
in Sec.~\ref{sec:theory} (in connection with Eq.~\ref{eq:D_P_liv_num})
that when both the LIV parameters $\eem$ and $\eet$ 
are present together, they may cancel out the impact
of each other to a significant extent. In fact, 
we corroborate this reality while showing DUNE's
octant discovery potential in our results section.

\section{Our results}
\label{sec:results}

In this section, we present our sensitivity results.
We start the discussion by showing the octant 
discovery potential of DUNE as a function of true
values of the standard CP phase $\delta$ for 
both SI and SI+LIV schemes.

\subsection{Octant discovery potential as a function of true $\dcp$}
\label{ssec:chisq_octant_delta}

\begin{figure}[h!]
\centering
\includegraphics[scale=0.6]{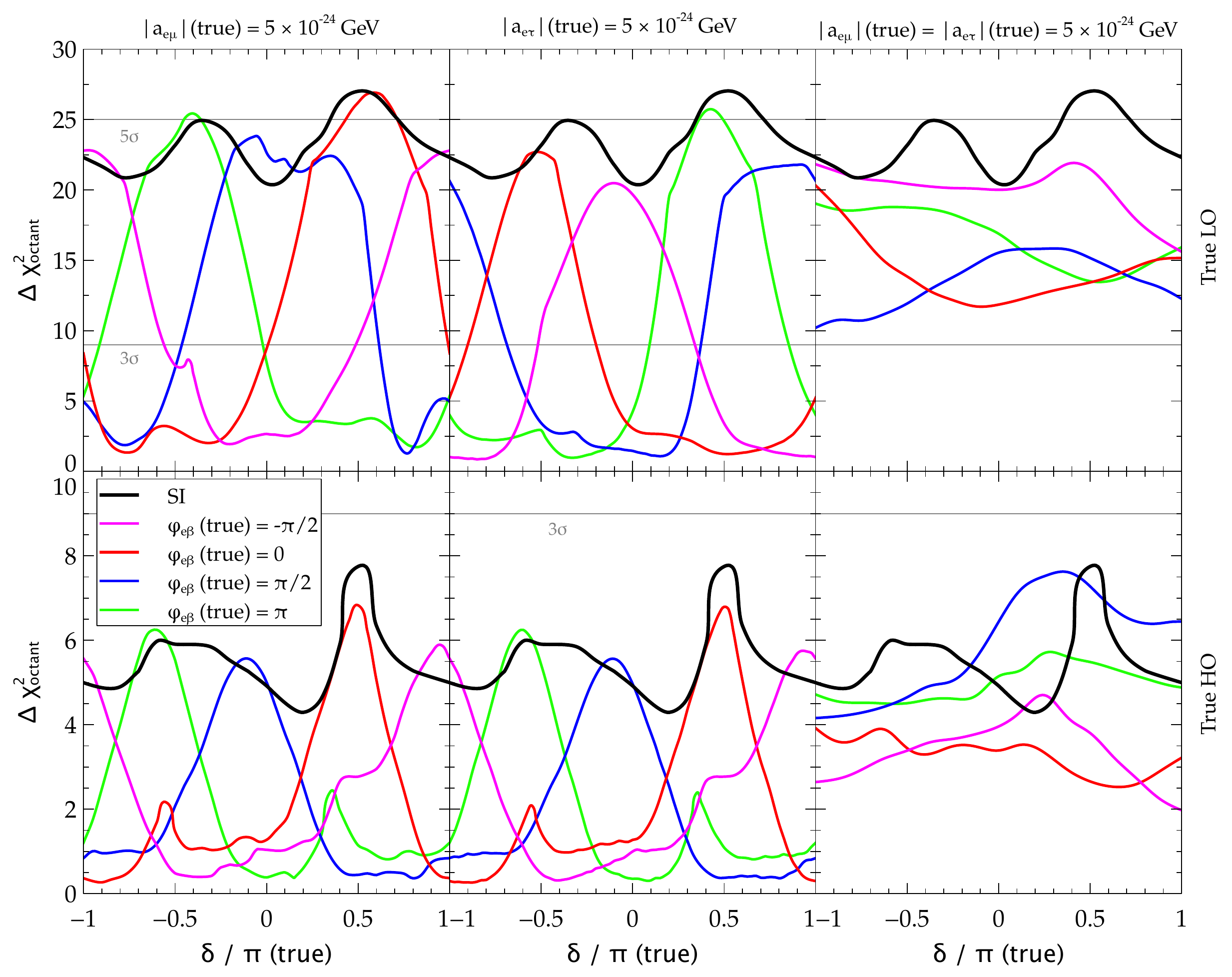}
\caption{\footnotesize{Discovery potential
of the true octant as a function of true $\delta$. 
In top (bottom) panels, we assume NO-LO 
(NO-HO) as the true choice with 
$\theta_{23}^{\textrm{true}} = 42.3^{\circ}$ 
($47.7^{\circ}$) as benchmark value for 
LO (HO) case. The left (middle) panels
are for the individual LIV parameter
$a_{e\mu}$ ($a_{e\tau}$), while the 
right panels deal with the case when both 
the LIV parameters are present simultaneously.
In each panel, the black 
curve shows the result for the SI case, 
while the four colored lines depict the 
sensitivity for the SI+LIV scheme 
considering four different true values of 
$\varphi_{e\mu}$ (left panel), 
$\varphi_{e\tau}$ (middle panel),
and $\varphi_{e\mu}$, $\varphi_{e\tau}$ 
(right panel), as mentioned in the legends.
In all the panels, the strength of the true 
LIV parameter is assumed to be 
$5 \times 10^{-24}$ GeV. 
See text for marginalization and other details.
Note that the y-axis ranges are different in
top and bottom panels.}}
\label{fig:chisq_del}
\end{figure}

Fig.~\ref{fig:chisq_del} exhibits the sensitivity
(in terms of $\Delta \chi^{2}$) for excluding 
the wrong octant as a function of the true 
values of the standard Dirac CP phase $\delta$. 
The top (bottom) panels show the results
assuming NO-LO (NO-HO) as the true choice 
with $\theta_{23}^{\textrm{true}} = 42.3^{\circ}$ 
($47.7^{\circ}$) for LO (HO) case. 
The left (middle) panels consider the individual 
LIV parameter $a_{e\mu}$ ($a_{e\tau}$), 
while the right panels depict the case when 
both the LIV parameters are present together
in the analysis. The black line in each panel
represents the octant sensitivity for the SI 
case, while the four colored lines show the 
octant sensitivity for the SI+LIV framework
considering four different true values of 
$\varphi_{e\mu}$ (left panel), 
$\varphi_{e\tau}$ (middle panel),
and $\varphi_{e\mu}$, $\varphi_{e\tau}$ 
(right panel), as mentioned in the legends.
In all the panels, the strength of the true LIV 
parameter is taken to be $5 \times 10^{-24}$ 
GeV. As discussed in Sec.~\ref{sec:simulation},
in the test dataset, $\theta_{23}^{\textrm{test}}$
has been marginalized over its all possible 
values in the wrong/opposite octant including 
the maximal value ($45^{\circ}$). In the SI case, 
we perform the marginalization over 
$\delta^{\textrm{test}}$ in its entire range 
of $[-\pi, \pi]$, while in the SI+LIV scheme,
we marginalize over both $\delta^{\textrm{test}}$
and $\varphi_{e\beta}^{\textrm{test}}$
(where $\beta$ can be $\mu$ or $\tau$) 
in their full parameter space of $[-\pi, \pi]$.
We also marginalize over $|a_{e\beta}|^{\textrm{test}}$ 
in the range of $5 \times 10^{-25}$ GeV 
to $5 \times 10^{-23}$ GeV as mentioned 
in Sec.~\ref{sec:simulation}. We note the 
following features from Fig.~\ref{fig:chisq_del}.
\begin{itemize}

\item
For true LO (see top panels), the octant sensitivity 
for the SI case lies roughly between 4.5$\sigma$ 
to 5.2$\sigma$ depending on the true value of 
$\delta$. In presence of $|\eem|$ (true) = $5 \times 10^{-24}$ 
GeV, the sensitivity can be as low as $\sim 1\sigma$ 
depending on the value of true $\delta$ and true $\eemp$ 
(see top left panel). Similar degradation in the sensitivity 
is also observed in presence of $\eet$ (see top middle panel). 
Since the standard CP phase $\delta$, as well as the LIV 
CP phases $\eemp$ and $\eetp$ are still undetermined, 
such spoiling of octant sensitivities is very much possible.

\item
For true HO (see bottom panels), the octant sensitivity 
for the SI case is relatively lower (approximately 
2.2$\sigma$ to 2.7$\sigma$). Here, in presence 
of $\eem$ or $\eet$, one can observe similar reduction 
in the sensitivity as we notice in the true LO case. 
The sensitivity may decrease to very small values
($\lesssim$ 1$\sigma$) for many choices of true values
of $\delta$ and $\varphi_{{e\mu}/{e\tau}}$ (see bottom 
left and bottom middle panels).

\item
In the right panels, in presence of both the LIV parameters 
$\eem$ and $\eet$ with the same magnitude, the worsening 
in the octant sensitivity is significantly less than what we observe
for the single LIV parameter case. For true LO (HO), the sensitivity 
does not go below $\sim 3.2\sigma$ ($1.5 \sigma$). As we have 
discussed in sections \ref{sec:theory} and \ref{sec:prob_event}, 
this is due to the fact that $\eem$ and $\eet$ effectively nullify 
the impact of each other to a significant extent (due to the presence 
of a relative sign between the $\Delta P_{\mu e}(\eem)$ 
and $\Delta P_{\mu e}(\eet)$ terms in Eq.~\ref{eq:D_P_liv_num}). 
This very interesting and counterintuitive impact of LIV on octant 
sensitivity is discussed for the first time in the present work.

\item
The $\Delta \chi^{2}$ curves in presence of LIV have 
prominent peaks, which are more apparent for true LO. 
In presence of the LIV parameter $\eem$, the choice 
of CP-conserving true values of $\eemp = 0, \pi$ (\ie, red 
and green curves, respectively) roughly produces peaks 
around maximal CP-violating true values of $\delta \simeq 
\pi/2, -\pi/2$, respectively. Conversely, maximal CP-violating 
true values of $\eemp = \pi/2, -\pi/2$ (\ie, blue and magenta 
curves, respectively) approximately produce peaks near 
CP-conserving true values of $\delta \simeq 0, \pm \pi$, 
respectively. This trend can also be observed in presence 
of $\eet$ with the location of the peaks interchanged. 
In presence of both $\eem$ and $\eet$, such a prominent 
feature is not noticed. 

\end{itemize}

\subsection{Octant discovery potential in [$\sin^2\theta_{23}$ -- $\dcp$] (true) plane}
\label{ssec:chisq_octant_delta_th23}

\begin{figure}[h!]
\centering
\includegraphics[scale=0.75]{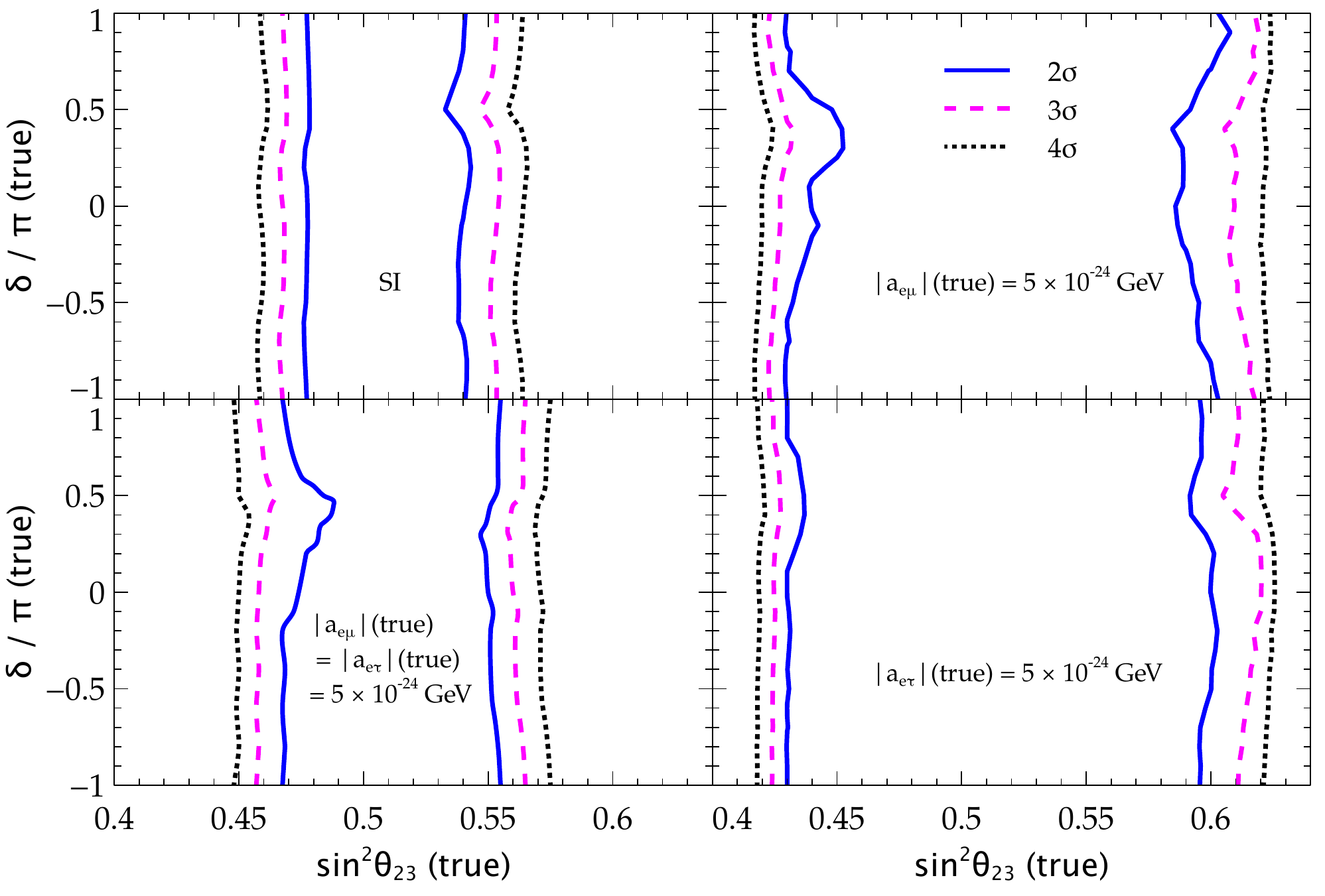}
\caption{\footnotesize{Octant discovery 
potential in [$\sin^2\theta_{23}$ -- $\dcp$] 
(true) plane at 2$\sigma$ (solid blue curves), 
3$\sigma$ (dashed magenta curves), and 
4$\sigma$ (dotted black curves) confidence
levels (1 d.o.f.) assuming NO both in data 
and theory. We consider four different 
scenarios: the SI case (top left panel), 
the SI+LIV case with $a_{e\mu}$ 
(top right panel), the SI+LIV scheme 
with $a_{e\tau}$ (bottom right panel), 
and the SI+LIV framework with both 
$a_{e\mu}$ and $a_{e\tau}$ present 
together (bottom left panel). In all the 
panels, the strength of the true LIV 
parameter is assumed to be 
$5 \times 10^{-24}$ GeV. See text for 
marginalization and other details.}}
\label{fig:true_th23_del-1}
\end{figure}

\begin{figure}[h!]
\centering
\includegraphics[scale=0.75]{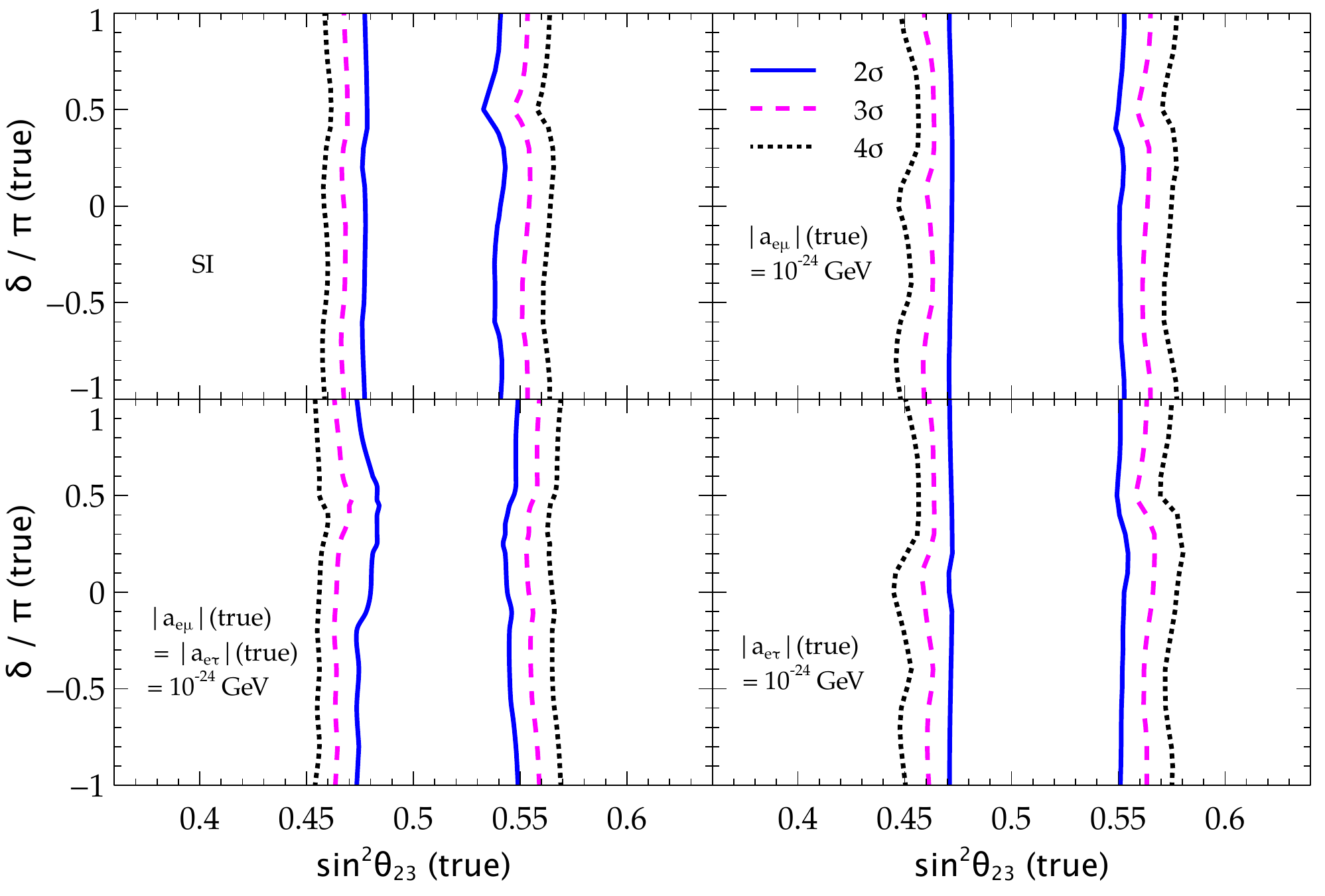}
\caption{\footnotesize{Octant discovery 
potential in [$\sin^2\theta_{23}$ -- $\dcp$] 
(true) plane at 2$\sigma$ (solid blue curves), 
3$\sigma$ (dashed magenta curves), and 
4$\sigma$ (dotted black curves) confidence
levels (1 d.o.f.) assuming NO both in data 
and theory. We consider four different 
scenarios: the SI case (top left panel), 
the SI+LIV case with $a_{e\mu}$ 
(top right panel), the SI+LIV scheme 
with $a_{e\tau}$ (bottom right panel), 
and the SI+LIV framework with both 
$a_{e\mu}$ and $a_{e\tau}$ present 
together (bottom left panel). In all the 
panels, the strength of the true LIV 
parameter is assumed to be 
$10^{-24}$ GeV. See text for 
marginalization and other details.}}
\label{fig:true_th23_del-2}
\end{figure}

\begin{figure}[h!]
\centering
\includegraphics[scale=0.75]{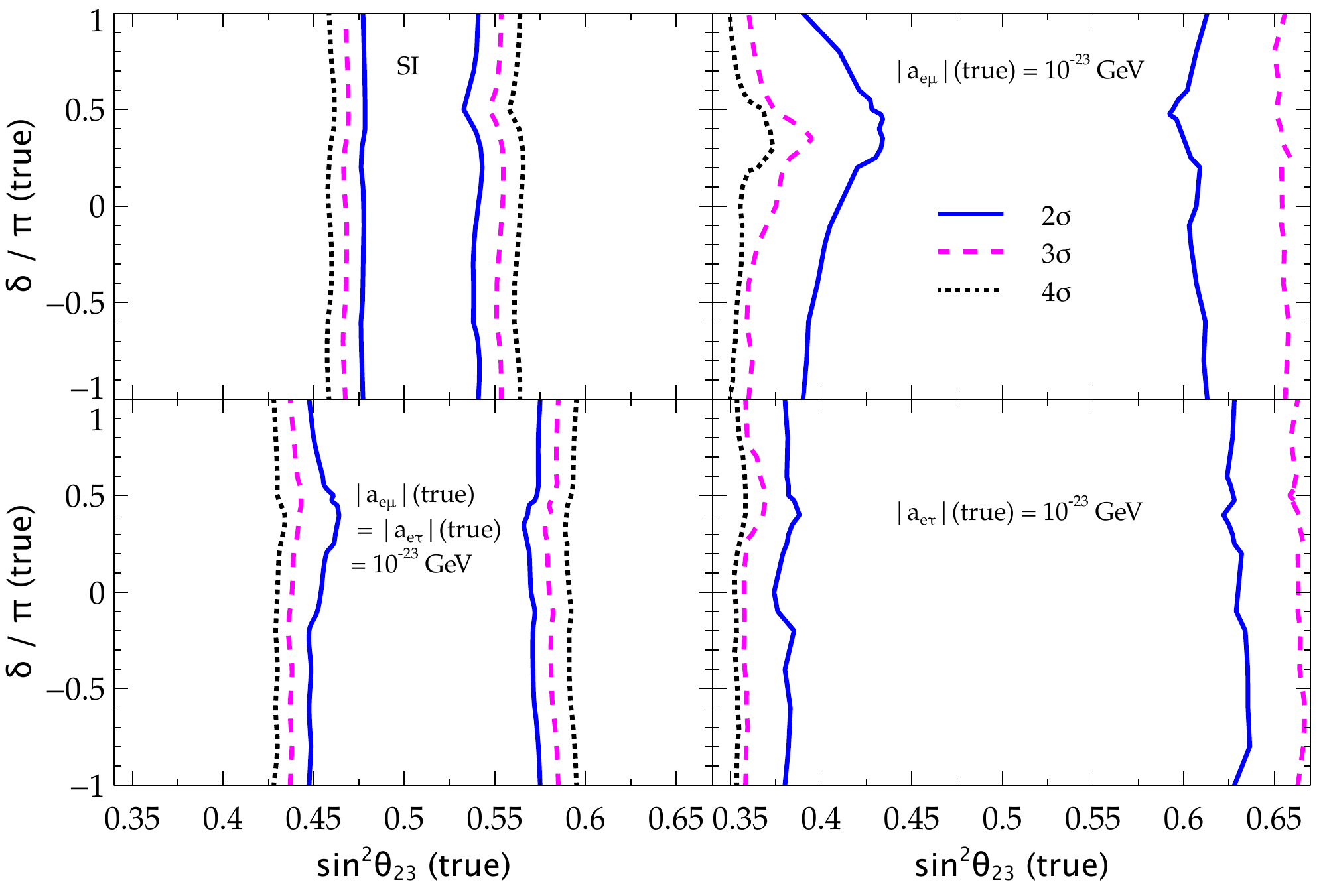}
\caption{\footnotesize{Octant discovery 
potential in [$\sin^2\theta_{23}$ -- $\dcp$] 
(true) plane at 2$\sigma$ (solid blue curves), 
3$\sigma$ (dashed magenta curves), and 
4$\sigma$ (dotted black curves) confidence
levels (1 d.o.f.) assuming NO both in data 
and theory. We consider four different 
scenarios: the SI case (top left panel), 
the SI+LIV case with $a_{e\mu}$ 
(top right panel), the SI+LIV scheme 
with $a_{e\tau}$ (bottom right panel), 
and the SI+LIV framework with both 
$a_{e\mu}$ and $a_{e\tau}$ present 
together (bottom left panel). In all the 
panels, the strength of the true LIV 
parameter is assumed to be $10^{-23}$ 
GeV. See text for marginalization 
and other details.}}
\label{fig:true_th23_del-3}
\end{figure}

The exact value of $\sin^{2}\theta_{23}$ 
is still to be determined. Therefore, we
consider all the allowed values of 
$\sin^{2}\theta_{23}$ and portray in 
Fig.~\ref{fig:true_th23_del-1}, the discovery 
potential of the true octant in the parameter
space of true $\sin^{2}\theta_{23}$ -- 
true $\delta$ at 2$\sigma$ (solid blue curves), 
3$\sigma$ (dashed magenta curves), and 
4$\sigma$ (dotted black curves) confidence
levels (1 d.o.f.) such that 
$\sigma = \sqrt{\Delta\chi^{2}}$.
We explore four different schemes: 
the SI case (top left panel), 
the SI+LIV case with $a_{e\mu}$ 
(top right panel), the SI+LIV scheme 
with $a_{e\tau}$ (bottom right panel), 
and the SI+LIV framework with both 
$a_{e\mu}$ and $a_{e\tau}$ present 
together in the simulation (bottom left panel). 
In all the panels, the strength of the 
true LIV parameter is assumed to be 
$5 \times 10^{-24}$ GeV.
Here, we assume NO both in data and
theory. In case of SI, we calculate the
$\Delta \chi^{2}$ by marginalizing over
test $\delta$ ($\in [-\pi, \pi]$) and test
$\theta_{23}$ over the wrong 
octant\footnote{For instance, for every 
true $\theta_{23}$ in the LO 
(\ie, $\theta_{23}^{\text{true}} 
< 45^{\circ}$), the test $\theta_{23}$ is 
marginalized over the entire allowed 
range in HO (\ie, $\theta_{23}^{\text{test}}  
\in [45^{\circ}, 50.7^{\circ}]$). 
Similarly, for every
true $\theta_{23}$ in the HO 
(\ie, $\theta_{23}^{\text{true}}
> 45^{\circ}$), the test $\theta_{23}$ 
is marginalized over the entire allowed 
range in LO (\ie, $\theta_{23}^{\text{test}}
\in [41.8^{\circ}, 45^{\circ}]$).}. 
In the SI+LIV schemes, we additionally
marginalize over $|a_{e\mu}|^{\textrm{test}}$,
the true and test values of the new dynamical 
CP phase $\varphi_{e\mu}$ (top left panel)
and $|a_{e\tau}|^{\textrm{test}}$, 
$\varphi_{e\tau}^{\textrm{true}}$,
$\varphi_{e\tau}^{\textrm{test}}$ 
(bottom right panel). In bottom left
panel, we marginalize over 
$|a_{e\mu}|^{\textrm{test}}$,
$|a_{e\tau}|^{\textrm{test}}$,
and the true and test values of 
the additional CP phases $\varphi_{e\mu}$ 
and $\varphi_{e\tau}$ in their entire 
allowed range of $[-\pi, \pi]$. 
It is apparent that in presence 
of the individual LIV parameter 
$a_{e\mu}$ (top right panel) or 
$a_{e\tau}$ (bottom right panel),
the sensitivity towards the octant
of $\theta_{23}$ gets reduced 
considerably as compared to the
SI case (top left panel). In such 
cases, the octant of $\theta_{23}$
can only be resolved at $3\sigma$ 
confidence level if the true value of 
$\theta_{23}$ turns out to be at least
$5^{\circ}$ to $7^{\circ}$ away from 
maximal mixing ($45^{\circ}$) for any 
choices of $\delta$ and 
$\varphi_{{e\mu}/{e\tau}}$. 
When both the LIV parameters 
$a_{e\mu}$ and $a_{e\tau}$ 
are present together in the simulation 
(see bottom left panel), they cancel 
their effect to a large extent, and we 
see a very slight deterioration in the 
octant sensitivity as compared to the
SI case. We observe similar feature
in the previous section as well.

In Fig.~\ref{fig:true_th23_del-2}
(Fig.~\ref{fig:true_th23_del-3}),
we portray the same assuming
the strength of the true LIV parameter 
to be $10^{-24}$ GeV ($10^{-23}$ GeV).
It is evident from Eq.~\ref{eq:p_liv} that
the impact of the LIV in $\nu_\mu$ to 
$\nu_e$ transition channel is proportional to the 
strength of the LIV parameter $|a_{e\beta}|$ 
(where $\beta$ can be $\mu$ or $\tau$).
For this reason, as we decrease the
strength of the LIV parameter in data 
from $5 \times 10^{-24}$ GeV 
(see Fig.~\ref{fig:true_th23_del-1})
to $10^{-24}$ GeV (see Fig.~\ref{fig:true_th23_del-2}),
we see a very minimal deterioration
in the octant discovery potential in 
the SI+LIV case as compared to the
SI case. On the contrary, if we increase
the strength of the true LIV parameter from
$5 \times 10^{-24}$ GeV 
(see Fig.~\ref{fig:true_th23_del-1})
to $10^{-23}$ GeV (see Fig.~\ref{fig:true_th23_del-3}),
we observe a huge deterioration in the
octant discovery potential in the SI+LIV 
scheme as compared to the SI scheme.
In Fig.~\ref{fig:true_th23_del-3}, when
we consider one LIV parameter at-a-time
(top and bottom right panels), the worsening
in the octant discovery potential is so large 
that even at 2$\sigma$ confidence level,
the true values of $\sin^2\theta_{23}$ for
which DUNE would be able to resolve the
octant, almost go beyond the present 
3$\sigma$ allowed range.

\subsection{Octant discovery potential as a function of LIV strength}
\label{ssec:chisq_octant_eps_th23}

\begin{figure}[h!]
\centering
\includegraphics[scale=0.62]{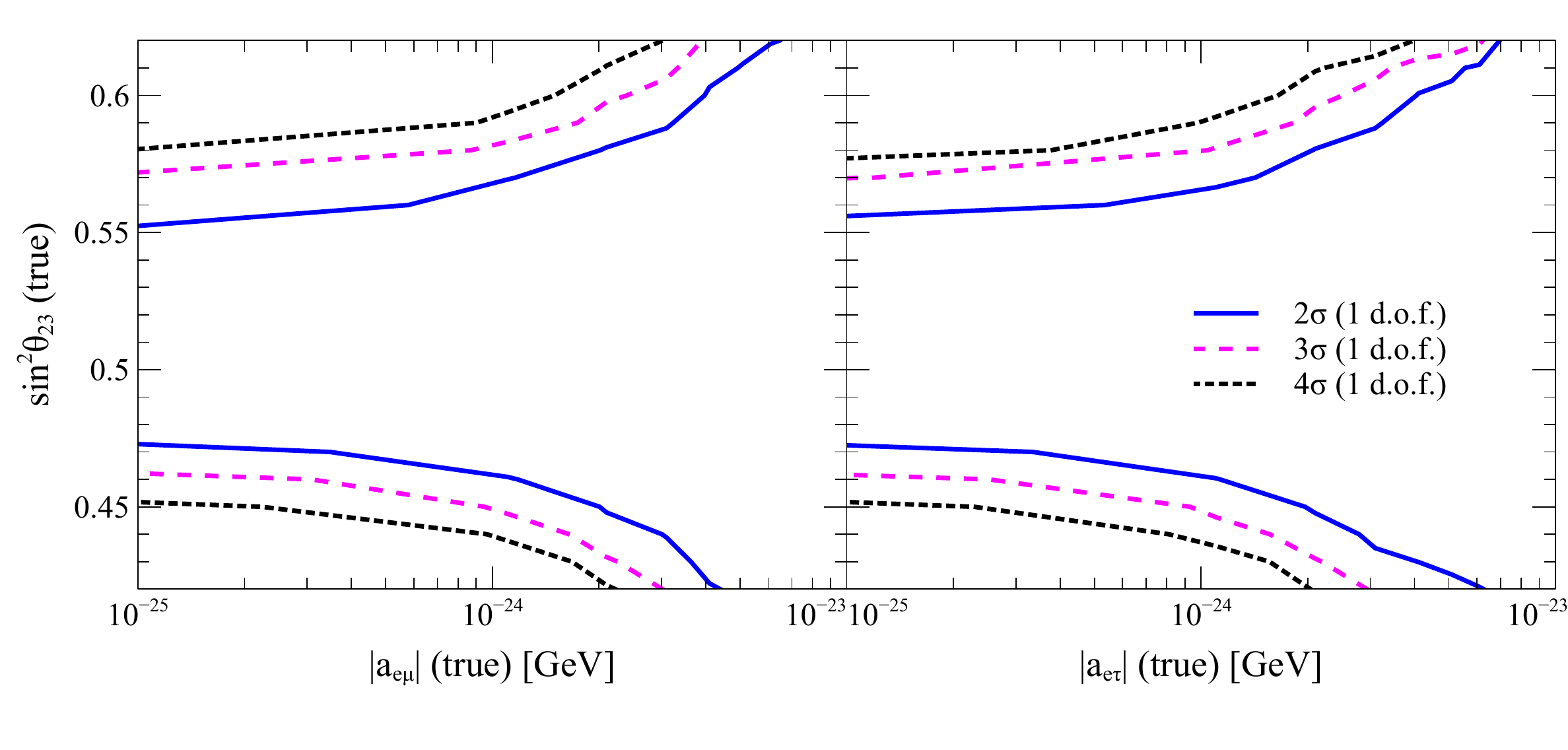}
\caption{\footnotesize{Deterioration 
of the $\theta_{23}$ octant discovery 
potential as a function of the strength 
of the true LIV parameter $|a_{e\mu}|$ 
($|a_{e\tau}|$) in left (right) panel
assuming NO both in data and 
theory. In both the panels, we marginalize
over the test values of $\theta_{23}$
in the wrong octant and the standard 
CP phase $\delta$ (both true and test)
in its full range of $[-\pi, \pi]$. We marginalize 
over test choices of $|a_{e\mu}|$ 
($|a_{e\tau}|$) in the left (right) panel.
In addition to this, in the left (right)
panel, the true and test values of the 
new CP phase $\varphi_{e\mu}$
($\varphi_{e\tau}$) have been 
marginalized away. We show the 
results at three different confidence
levels (1 d.o.f.): 2$\sigma$ 
(solid blue curves), 3$\sigma$
(dashed magenta curves), and
4$\sigma$ (dotted black curves).}}
\label{fig:true_th23_eps}
\end{figure}

So far, we have shown our results for
few benchmark true values of the LIV
parameters $|a_{e\mu}|$ and $|a_{e\tau}|$.
Now, it would be quite interesting to see 
how the octant discovery potential gets 
modified if we vary the strength
of the LIV parameters $|a_{e\mu}|$ (true) 
and $|a_{e\tau}|$ (true). We present the result 
of this analysis in Fig.~\ref{fig:true_th23_eps}, 
which exhibits the discovery potential of the 
$\theta_{23}$ octant as a function of the 
strength of the true LIV parameter $|a_{e\mu}|$ 
($|a_{e\tau}|$) in left (right) panel assuming 
NO both in data and theory. In both the panels, 
we marginalize over the test values of 
$\theta_{23}$ in the wrong octant and 
the standard CP phase $\delta$ 
(both true and test) in its full range of 
$[-\pi, \pi]$. We marginalize over test
choices of $|a_{e\mu}|$ ($|a_{e\tau}|$)
in the left (right) panel. In addition to this, 
in the left (right) panel, the true and test 
values of the new CP phase $\varphi_{e\mu}$
($\varphi_{e\tau}$) have been marginalized 
away in the entire range of $[-\pi,\pi]$.
We show the results at three different 
confidence levels (1 d.o.f.): 2$\sigma$ 
(solid blue curves), 3$\sigma$ 
(dashed magenta curves), and
4$\sigma$ (dotted black curves).
It is clear that as the strength of the 
LIV parameter increases, the discovery 
potential of the true octant gets 
deteriorated gradually. We notice that
as the strength of the LIV parameters
approaches towards $10^{-25}$ GeV,
the sensitivities slowly get improved
and attain the values which we obtain
in the SI case. 

\subsection{Reconstruction of the CP phases}
\label{ssec:chisq_cp_recon}

\begin{figure}[h!]
\centering
\includegraphics[scale=0.59]{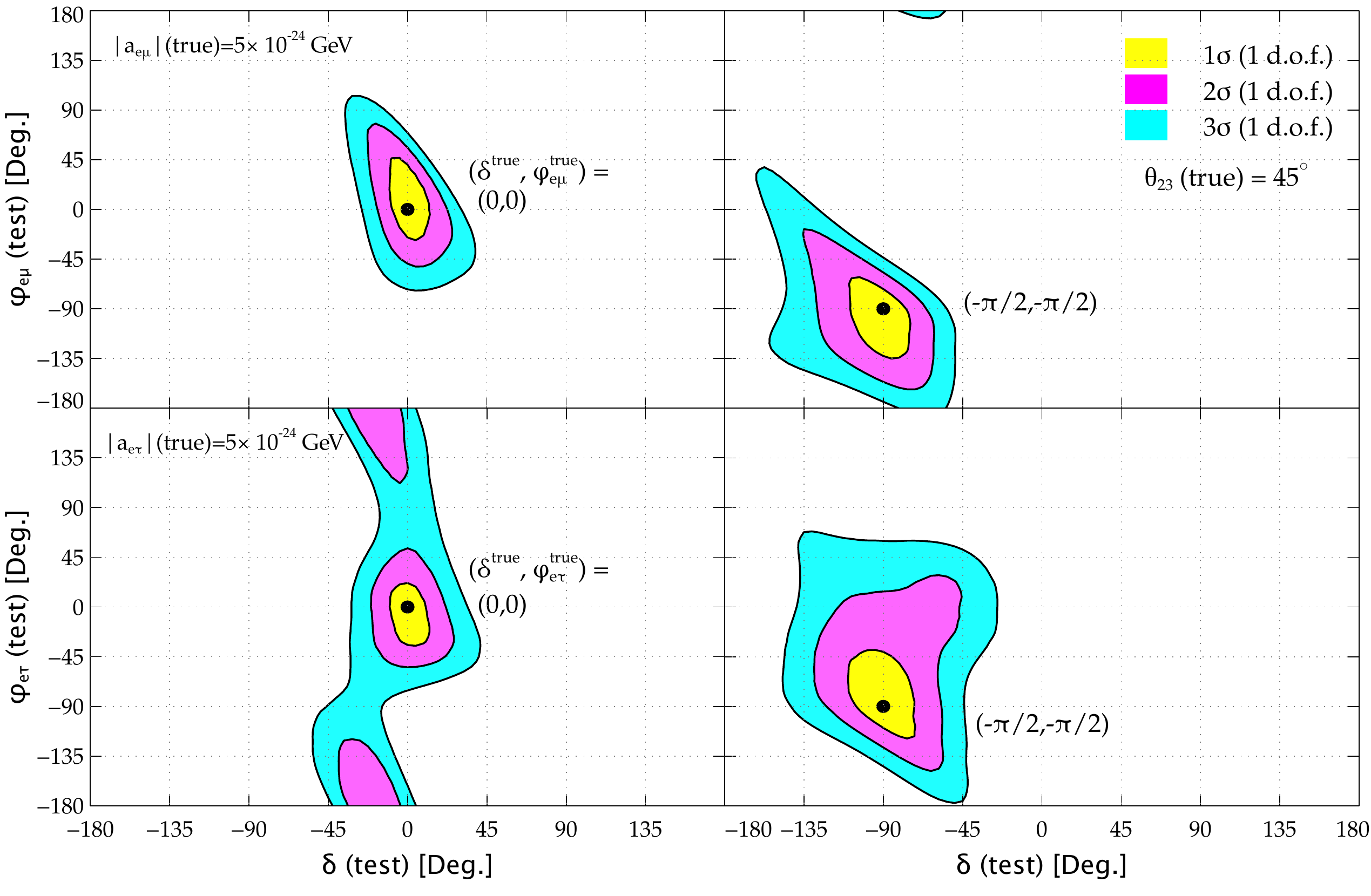}
\caption{\footnotesize{Reconstructed
regions for the two CP phases
$\delta$ and $\varphi_{e\mu}$
(top panels) at 1$\sigma$, 2$\sigma$,
and 3$\sigma$ (1 d.o.f.) confidence
levels assuming NO both in data 
and theory. The bottom panels show 
the same for the two CP phases
$\delta$ and $\varphi_{e\tau}$.
The two left (right) panels refer to 
the representative true values of 
the phases [0, 0] ([$- \pi/2, - \pi/2$]).
In all the panels, the strength of the 
true LIV parameter is assumed to be 
$5 \times 10^{-24}$ GeV, and we
marginalize over test choices of 
LIV parameter in the fit. We consider 
$\theta_{23}^{\text{true}} = 
45^{\circ}$ and marginalize
over $\theta_{23}^{\text{test}}$  
in the range $[41.8^{\circ}, 
50.7^{\circ}]$ in the fit.}} 
\label{fig:cp_recon_1}
\end{figure}

\begin{figure}[h!]
\centering
\includegraphics[scale=0.59]{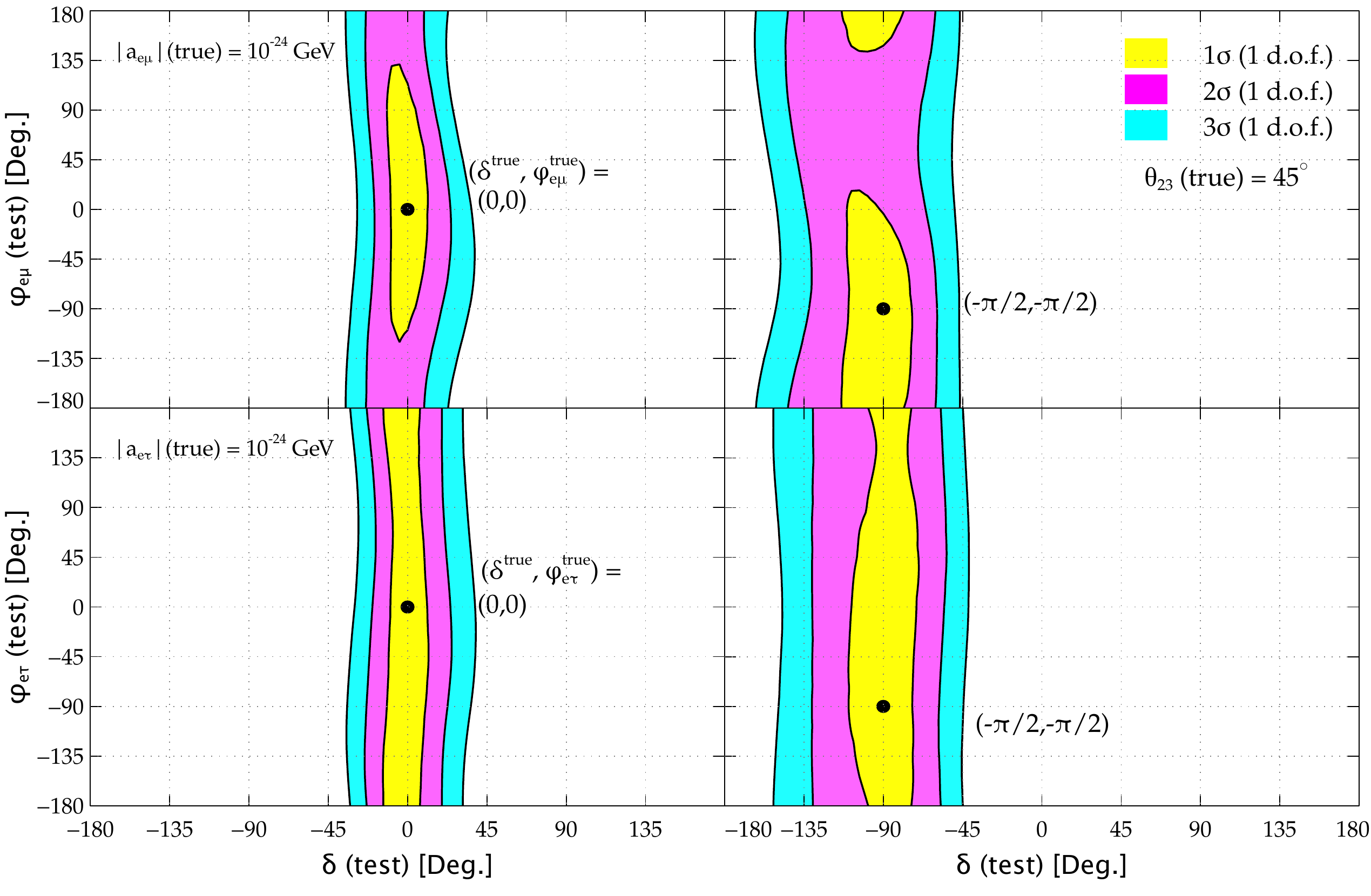}
\caption{\footnotesize{Reconstructed
regions for the two CP phases
$\delta$ and $\varphi_{e\mu}$
(top panels) at 1$\sigma$, 2$\sigma$,
and 3$\sigma$ (1 d.o.f.) confidence
levels assuming NO both in data 
and theory. The bottom panels show 
the same for the two CP phases
$\delta$ and $\varphi_{e\tau}$.
The two left (right) panels refer to 
the representative true values of 
the phases [0, 0] ([$- \pi/2, - \pi/2$]).
In all the panels, the strength of the 
true LIV parameter is assumed to be 
$10^{-24}$ GeV, and we
marginalize over test choices of 
LIV parameter in the fit.
We consider 
$\theta_{23}^{\text{true}} = 
45^{\circ}$ and marginalize
over $\theta_{23}^{\text{test}}$  
in the range $[41.8^{\circ}, 
50.7^{\circ}]$ in the fit.}} 
\label{fig:cp_recon_2}
\end{figure}

\begin{figure}[h!]
\centering
\includegraphics[scale=0.59]{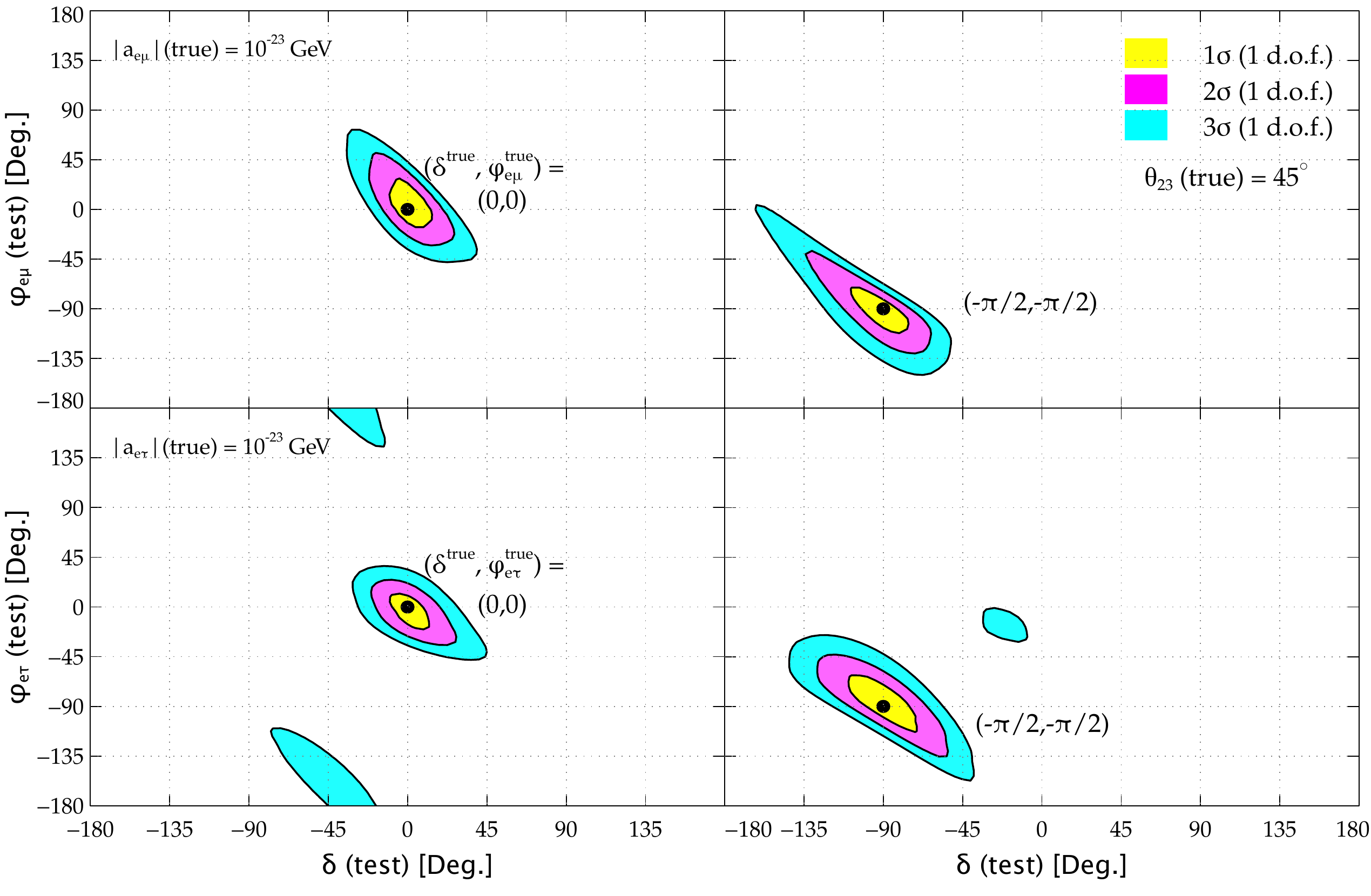}
\caption{\footnotesize{Reconstructed
regions for the two CP phases
$\delta$ and $\varphi_{e\mu}$
(top panels) at 1$\sigma$, 2$\sigma$,
and 3$\sigma$ (1 d.o.f.) confidence
levels assuming NO both in data 
and theory. The bottom panels show 
the same for the two CP phases
$\delta$ and $\varphi_{e\tau}$.
The two left (right) panels refer to 
the representative true values of 
the phases [0, 0] ([$- \pi/2, - \pi/2$]).
In all the panels, the strength of the 
true LIV parameter is assumed to be 
$10^{-23}$ GeV, and we marginalize 
over test choices of LIV parameter 
in the fit. We consider 
$\theta_{23}^{\text{true}} = 
45^{\circ}$ and marginalize
over $\theta_{23}^{\text{test}}$  
in the range $[41.8^{\circ}, 
50.7^{\circ}]$ in the fit.}} 
\label{fig:cp_recon_3}
\end{figure}

In the previous subsections,
we discuss in detail how the 
standard CP phase $\delta$
and the new dynamical CP
phase $\varphi_{e\mu}$/
$\varphi_{e\tau}$ would affect
the measurement of the octant 
of $\theta_{23}$ in DUNE. In 
this subsection, we explore the 
capability of DUNE in reconstructing
the true values of the CP phases
$\delta$ and $\varphi_{e\beta}$ 
(where $\beta$ can be $\mu$ 
or $\tau$). Fig.~\ref{fig:cp_recon_1}
shows the reconstructed regions 
for the two CP phases $\delta$ 
and $\varphi_{e\mu}$ (top panels) 
at 1$\sigma$, 2$\sigma$, and 
3$\sigma$ (1 d.o.f.) confidence
levels assuming NO both in data 
and theory. The bottom panels in
Fig.~\ref{fig:cp_recon_1} portray 
the same for the two CP phases
$\delta$ and $\varphi_{e\tau}$.
The two left panels refer to the
typical CP-conserving true values
of the phases [0, 0], while the two
right panels deal with the illustrative 
CP-violating true values of the 
phases [$- \pi/2, - \pi/2$]. 
In all the panels, the strength of the 
true LIV parameter is assumed to be 
$5 \times 10^{-24}$ GeV, and we
marginalize over test choices of 
LIV parameter in the fit. 
While generating the 
prospective data for DUNE, we 
consider the true value of 
$\theta_{23}$ to be $45^{\circ}$
and in the fit, we marginalize 
over the test values of $\theta_{23}$
in its 3$\sigma$ allowed range
of $41.8^{\circ}$ to $50.7^{\circ}$.
DUNE can measure the CP phases
$\delta$ and $\varphi_{e\mu}$
quite efficiently providing a unique 
reconstructed region around 
$\delta^{\text{true}} = 
\varphi_{e\mu}^{\text{true}} =
0^{\circ}$ at 3$\sigma$ confidence
level (see top left panel).
But in presence of $a_{e\tau}$ 
(see bottom left panel), the 
reconstruction becomes quite 
poor for $\varphi_{e\tau}$ at
2$\sigma$ and above.
For maximal CP-violating choices 
($- \pi/2, - \pi/2$) of the true CP 
phases (see top right and bottom
right panels), the reconstruction of
$\delta^{\text{true}}$ becomes
slightly worse as compared to 
the CP-conserving scenarios 
at 1$\sigma$ confidence level, 
while the reconstruction of 
$\varphi_{e\mu}^{\text{true}}$
and $\varphi_{e\tau}^{\text{true}}$
remains more or less same.
In Table~\ref{tab:cp_recon},
we mention the typical 
1$\sigma$ allowed ranges around 
the true values of the CP phases 
$\delta$ and $\varphi_{e\beta}$ 
(where $\beta$ can be $\mu$ 
or $\tau$). Here, we assume 
the strength of the true LIV parameters 
$|a_{e\mu}|$ and $|a_{e\tau}|$ 
to be $5 \times 10^{-24}$ GeV, 
and we marginalize over the 
test choices of the corresponding 
LIV parameters in the fit. We have 
also checked that the reconstruction 
of the CP phases becomes worse 
as $\theta_{23}$ deviates from 
the maximal mixing.

In Fig.~\ref{fig:cp_recon_2} and
Fig.~\ref{fig:cp_recon_3}, we show
the performance of DUNE in
reconstructing the true values of
the CP phases $\delta$ and 
$\varphi_{e\beta}$ (where $\beta$ 
can be $\mu$ or $\tau$) considering
the strength of the true LIV parameter
to be $10^{-24}$ GeV and 
$10^{-23}$ GeV, respectively.
As expected, once we decrease
the strength of the true LIV parameter
$|a_{e\mu}|$/$|a_{e\tau}|$ from 
$5 \times 10^{-24}$ GeV 
(see Fig.~\ref{fig:cp_recon_1}) to 
$10^{-24}$ GeV (see Fig.~\ref{fig:cp_recon_2}),
the reconstruction of the new 
dynamical CP phases 
$\varphi_{e\mu}$/$\varphi_{e\tau}$ 
becomes quite impossible, and
almost 75\% values of $\varphi_{e\mu}$ 
and all the values of $\varphi_{e\tau}$
get allowed even at 1$\sigma$
confidence level.  As far as the
reconstruction of the standard 
CP phase $\delta$ is concerned,
the allowed regions for $\delta$
remain almost the same 
in Figs.~\ref{fig:cp_recon_1}
and \ref{fig:cp_recon_2}.
When we increase the strength 
of the true LIV parameter 
$|a_{e\mu}|$/$|a_{e\tau}|$ 
in data from $5 \times 10^{-24}$ GeV 
(see Fig.~\ref{fig:cp_recon_1}) 
to $10^{-23}$ GeV 
(see Fig.~\ref{fig:cp_recon_3}),
the reconstruction of the new
dynamical CP phases 
$\varphi_{e\mu}$/$\varphi_{e\tau}$ 
gets improved significantly 
and the allowed regions for 
$\delta$ remain almost
unaltered.  

\begin{table}[h!]
\centering
\begin{tabular}{|c|c|c|}
\hline
True values & $1\sigma$ range in $\delta^{\text{test}}$ (Deg.) & $1\sigma$ range in $\varphi^{\text{test}}$ (Deg.)\\ 
\hline\hline
$(\delta, \eemp) = (0,0)$ 
& $-8^{\circ} \lesssim \delta^{\text{test}} \lesssim 12^{\circ}$
& $ -25^{\circ} \lesssim \eemp^{\text{test}} \lesssim 38^{\circ}$ \\ \hline
$(\delta, \eetp) = (0,0)$
& $-9^{\circ} \lesssim \delta^{\text{test}} \lesssim 10^{\circ}$
& $-33^{\circ} \lesssim \eetp^{\text{test}} \lesssim 21^{\circ}$ \\ \hline
$(\delta, \eemp) = (-\pi/2,-\pi/2)$ 
& $-108^{\circ} \lesssim \delta^{\text{test}} \lesssim -78^{\circ}$
& $-131^{\circ} \lesssim \eemp^{\text{test}} \lesssim -74^{\circ}$ \\ \hline
$(\delta, \eetp) = (-\pi/2,-\pi/2)$ 
& $-105^{\circ} \lesssim \delta^{\text{test}} \lesssim -73^{\circ}$
& $-108^{\circ} \lesssim \eetp^{\text{test}} \lesssim -42^{\circ}$ \\ \hline
\end{tabular}
\caption{\footnotesize{The typical 
1$\sigma$ allowed ranges around 
the true values of the CP phases 
$\delta$ and $\varphi_{e\beta}$ 
(where $\beta$ can be $\mu$ 
or $\tau$). Here, we assume 
the strength of the true LIV 
parameters $|a_{e\mu}|$ and 
$|a_{e\tau}|$ to be $5 \times 
10^{-24}$ GeV, and we marginalize 
over the test choices of the 
corresponding LIV parameters 
in the fit.}}
\label{tab:cp_recon}
\end{table}

\section{Summary and conclusion}
\label{sec:summary}

We have a well-defined neutrino roadmap to resolve 
the remaining fundamental unknowns, in particular, 
the determination of neutrino mass ordering, the
clear demonstration of leptonic CP-violation (CPV),
and the precision measurement of the oscillation
parameters with the help of upcoming high-precision
long-baseline neutrino oscillation experiment DUNE.
This experiment will perform a rigorous test of the 
three-flavor oscillation framework and play an important 
role to test the existence of various new physics 
scenarios if they at all exist in Nature. One such 
new physics scenario is Lorentz Invariance Violation (LIV).

In this paper, we study the impact of LIV in determining 
the octant of $\theta_{23}$ and in reconstructing the CP
phases considering the DUNE as a case study. 
We discuss in detail how the two most relevant 
CPT-violating LIV parameters
$a_{e\mu}$ ($\equiv |a_{e\mu}|e^{i\varphi_{e\mu}}$)
and $a_{e\tau}$ ($\equiv |a_{e\tau}|e^{i\varphi_{e\tau}}$)
affect neutrino and antineutrino appearance probabilities. 
These LIV parameters ($a_{e\mu}$ or $a_{e\tau}$) introduce 
an additional interference term in $\nu_{\mu} \to \nu_{e}$ 
and $\bar\nu_{\mu} \to \bar\nu_{e}$ oscillation channels. 
This new interference term depends on both the standard 
CP phase $\delta$ and the new dynamical CP phase 
$\varphi_{e\mu}$/$\varphi_{e\tau}$. This term gets 
summed up with the well-known interference term 
related to the standard CP phase $\delta$ and gives 
rise to new degeneracies among $\theta_{23}$, $\delta$, 
and $\varphi$. These new degeneracies spoil the
measurement of octant of $\theta_{23}$. We show 
that for values of the LIV parameter (taken one at-a-time) 
as small as $|a_{e\mu}| = |a_{e\tau}| = 5 \times 10^{-24}$ 
GeV, the octant discovery potential of DUNE deteriorates
considerably for unfavorable combinations of the two CP
phases $\delta$ and $\varphi_{e\mu}$/$\varphi_{e\tau}$.
DUNE can only resolve the octant ambiguity of $\theta_{23}$
at $3\sigma$ confidence level for any choices of $\delta$ 
and $\varphi$ if $\theta_{23}$ turns out to be at least 
$5^{\circ}$ to $7^{\circ}$ away from maximal mixing.
We also perform the analysis considering both the LIV
parameters $a_{e\mu}$ and $a_{e\tau}$ together and 
observe for the first time that they largely nullify the effect 
of each other due to the apparent relative sign between the 
$a_{e\mu}$-term and the $a_{e\tau}$-term in the probability 
expressions. For this reason, DUNE can retrieve its octant 
resolution capability if both the LIV parameters $a_{e\mu}$ 
and $a_{e\tau}$ are present together in the analysis. 
We also study how the deterioration of the $\theta_{23}$
octant discovery potential varies with the magnitude of
the LIV parameters in the [$\sin^2\theta_{23}$ (true) 
-- $|a_{e\beta}|$ (true)] plane where $\beta$ can be 
$\mu$ or $\tau$. We also address how well DUNE 
can reconstruct the standard CP phase $\delta$
and the new dynamical CP phase $\varphi_{e\mu}$/$\varphi_{e\tau}$. 
Our analysis reveal that the typical $1\sigma$ uncertainty 
on $\delta$ is $10^{\circ}$ to $15^{\circ}$ and the same 
on $\varphi_{e\mu}$/$\varphi_{e\tau}$ is $25^{\circ}$ 
to $30^{\circ}$. So, at the end, we conclude that a
small amount of Lorentz Invariance Violation (LIV)
may affect the measurements of octant of 2-3 mixing
angle and the CP phases at DUNE and we hope that 
our present study will be a valuable addition to the 
landscape of new physics scenarios beyond the standard
three-neutrino oscillation paradigm which can be probed 
using the DUNE facility.

\section{Acknowledgements}
S.K.A. is supported by the DST/INSPIRE Research Grant [IFA-PH-12] 
from the Department of Science and Technology (DST), India. 
S.K.A. and M.M. acknowledge the financial support from the 
Indian National Science Academy (INSA) Young Scientist Project 
[INSA/SP/YSP/144/2017/1578]. 

\begin{appendix}

\section{Octant discovery potential with LIV in data only}
\label{sec:LIV-in-data-not-in-fit}

\begin{figure}[h!]
\centering
\includegraphics[scale=0.6]{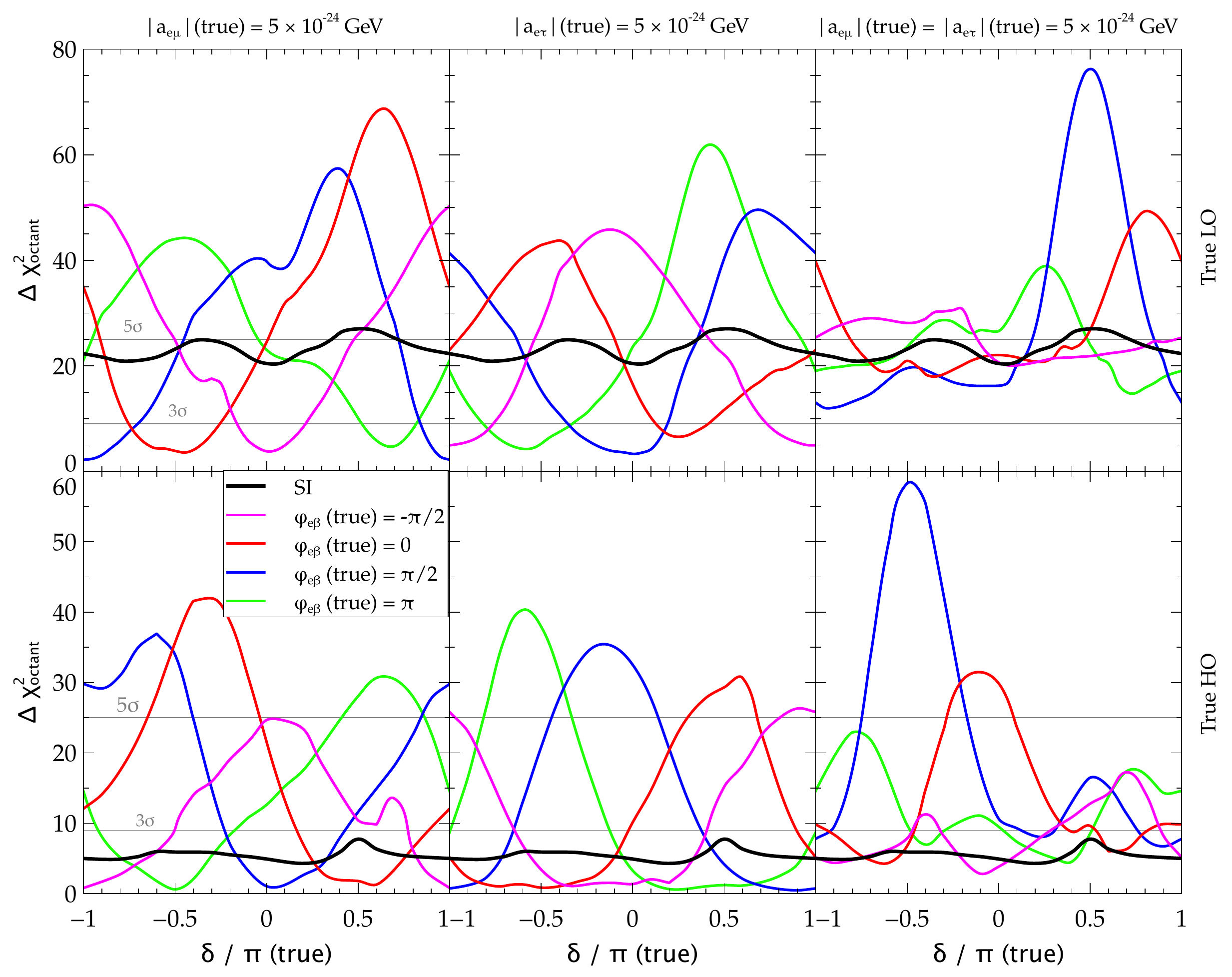}
\caption{\footnotesize{Octant discovery potential
as a function of true $\delta$. In top (bottom) panels, 
we assume NO-LO (NO-HO) as the true choice with 
$\theta_{23}^{\textrm{true}} = 42.3^{\circ}$ 
($47.7^{\circ}$) as benchmark value for LO (HO) case. 
In all the panels, the data are generated assuming
the strength of the true LIV parameter to be 
$5 \times 10^{-24}$ GeV, and we do not consider
the LIV parameters in the fit (theory). The left (middle) 
panels are for the individual LIV parameter
$a_{e\mu}$ ($a_{e\tau}$), while the 
right panels deal with the case when both 
the LIV parameters are present simultaneously.
In each panel, the black curve shows the result 
for the SI case, while the four colored lines depict 
the sensitivity for the SI+LIV scheme 
considering four different true values of 
$\varphi_{e\mu}$ (left panel), 
$\varphi_{e\tau}$ (middle panel),
and $\varphi_{e\mu}$, $\varphi_{e\tau}$ 
(right panel), as mentioned in the legends.
We marginalize over $\theta_{23}^{\textrm{test}}$
in the wrong/opposite octant including the 
maximal value ($45^{\circ}$) and $\delta^{\textrm{test}}$
in its entire range of $[-\pi, \pi]$. Note that the y-axis 
ranges are different in top and bottom panels.}}
\label{fig:chisq_LIV_data_only}
\end{figure}

Here, we explore the octant discovery potential 
of DUNE assuming the presence of LIV in data,
but not in fit (theory). Fig.~\ref{fig:chisq_LIV_data_only}
shows the octant discovery potential of DUNE
as a function of true $\delta$. In top (bottom) panels, 
we assume NO-LO (NO-HO) as the true choice with 
$\theta_{23}^{\textrm{true}} = 42.3^{\circ}$ 
($47.7^{\circ}$) as benchmark value for LO (HO) case. 
In all the panels, the data are generated assuming
the strength of the true LIV parameter to be 
$5 \times 10^{-24}$ GeV, and we do not consider
the LIV parameters in the fit (theory). The left (middle) 
panels are for the individual LIV parameter
$a_{e\mu}$ ($a_{e\tau}$), while the 
right panels deal with the case when both 
the LIV parameters are present simultaneously.
In each panel, the black curve shows the result 
for the SI case, while the four colored lines depict 
the sensitivity for the SI+LIV scheme 
considering four different true values of 
$\varphi_{e\mu}$ (left panel), 
$\varphi_{e\tau}$ (middle panel),
and $\varphi_{e\mu}$, $\varphi_{e\tau}$ 
(right panel), as mentioned in the legends.
We marginalize over $\theta_{23}^{\textrm{test}}$
in the wrong/opposite octant including the 
maximal value ($45^{\circ}$) and $\delta^{\textrm{test}}$
in its entire range of $[-\pi, \pi]$. Note that the y-axis 
ranges are different in top and bottom panels.
Since we introduce the effect of LIV in data, but
not in theory, then, due to this mismatch in data
and theory, we obtain additional contribution 
to the $\Delta \chi^{2}$ on top of the contribution 
due to the choices of opposite octants in data
and theory. It happens for most of the favorable
combinations of $\delta^{\textrm{true}}$ and 
$\varphi^{\textrm{true}}$ as can be seen from
Fig.~\ref{fig:chisq_LIV_data_only}. But there
are some unfavorable combinations of 
$\delta^{\textrm{true}}$ and $\varphi^{\textrm{true}}$
for which the sensitivities get deteriorated
substantially and the values of $\Delta \chi^{2}$
go below the SI case. It suggests that if we 
minimize the $\Delta \chi^{2}$ over the true
values of the standard CP phase $\delta$
and the new dynamical CP phase 
$\varphi_{e\mu}$/$\varphi_{e\tau}$, the
resulting $\Delta \chi^{2}$ attains a value
much lower than the value that we obtain
in the absence of LIV.



\end{appendix}

\bibliographystyle{JHEP}
\bibliography{reference}

\end{document}